\documentclass[sigconf]{acmart}

\usepackage{nicefrac}
\usepackage{siunitx}
\usepackage{array,framed}
\usepackage{booktabs}
\usepackage{
  color,
  float,
  epsfig,
  wrapfig,
  graphics,
  graphicx,
  sub caption
}
\PassOptionsToPackage{table,usenames,dvipsnames}{xcolor}
\usepackage{colortbl}  %
\usepackage{tcolorbox}
\usepackage{textcomp}
\usepackage{algorithm2e}
\usepackage{algpseudocode}
\usepackage{graphics}
\usepackage{censor}
\usepackage{xparse} %
\usepackage[normalem]{ulem}
\useunder{\uline}{\ul}{}
\usepackage{xspace}
\usepackage{multirow}
\usepackage{csvsimple}
\usepackage{balance}
\usepackage{natbib}
\usepackage{flushend}
\usepackage{appendix}
\usepackage{enumitem}

\usepackage{siunitx}
\usepackage{geometry}

\usepackage{
  tikz,
  pgfplots,
  pgfplotstable
}
\usepackage{hyperref}

\newcommand\PartCircle[2]{
  \begin{tikzpicture}
    \draw[fill=black, line width=0.8mm] (0,0) circle (0.1cm);
    \fill[white] (0,0) -- (90:0.1cm) arc[start angle=90, end angle=90-360*#1/#2, radius=0.1cm] -- cycle;
  \end{tikzpicture}
}

\usepackage{ulem}

\newtheorem{game}{Game}
\newtheorem{definition}{Definition}[section]

\usepackage{cleveref} %

\usepackage{listings}
\lstdefinestyle{mystyle}{
    language=Python,
    basicstyle=\ttfamily\small,
    keywordstyle=\color{blue}\bfseries,
    stringstyle=\color{red},
    commentstyle=\color{green}\itshape,
    frame=single,
    framerule=0.4pt,
    rulecolor=\color{black!20},
    backgroundcolor=\color{blue!10}
}

\AtBeginDocument{%
}

\copyrightyear{2024}
\acmYear{2024}
\setcopyright{rightsretained}
\acmConference[CCS '24]{Proceedings of the 2024 ACM SIGSAC Conference on Computer and Communications Security}{October 14--18, 2024}{Salt Lake City, UT, USA}
\acmBooktitle{Proceedings of the 2024 ACM SIGSAC Conference on Computer and Communications Security (CCS '24), October 14--18, 2024, Salt Lake City, UT, USA}
\acmDOI{10.1145/3658644.3690279}
\acmISBN{979-8-4007-0636-3/24/10}

\makeatletter \gdef\@copyrightpermission{
\begin{minipage}{0.3\columnwidth} \href{https://creativecommons.org/licenses/by/4.0/}{\includegraphics[width=0.90\textwidth]{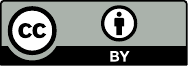}} \end{minipage}\hfill \begin{minipage}{0.7\columnwidth}
\href{https://creativecommons.org/licenses/by/4.0/}{This work is licensed under a Creative Commons Attribution International 4.0 License.}
\end{minipage}
\vspace{5pt} }
\makeatother

\begin{document}

\title{PreCurious: How Innocent Pre-Trained Language Models Turn into Privacy Traps}

\author{Ruixuan Liu}
\affiliation{%
  \institution{Emory University}
  \city{Atlanta}
  \country{USA}
}
\email{ruixuan.liu2@emory.edu}

\author{Tianhao Wang}
\affiliation{%
  \institution{University of Virginia}
  \city{Charlottesville}
  \country{USA}}
\email{tianhao@virginia.edu}

\author{Yang Cao}
\affiliation{%
  \institution{Tokyo Institute of Technology}
  \city{Tokyo}
  \country{Japan}
}
\email{cao@c.titech.ac.jp}

\author{Li Xiong}
\affiliation{%
 \institution{Emory University}
 \city{Atlanta}
 \country{USA}}
\email{lxiong@emory.edu}

\renewcommand{\shortauthors}{Ruixuan Liu, Tianhao Wang, Yang Cao, \& Li Xiong}

\begin{abstract}
The pre-training and fine-tuning paradigm has demonstrated its effectiveness and has become the standard approach for tailoring language models to various tasks.
Currently, community-based platforms offer easy access to various pre-trained models, as anyone can publish without strict validation processes.
However, a released pre-trained model can be a privacy trap for fine-tuning datasets if it is carefully designed.
In this work, we propose PreCurious framework to reveal the new attack surface where the attacker releases the pre-trained model and gets a black-box access to the final fine-tuned model. 
PreCurious aims to escalate the general privacy risk of both membership inference and data extraction on the fine-tuning dataset.
The key intuition behind PreCurious is to manipulate the memorization stage of the pre-trained model and guide fine-tuning with a seemingly legitimate configuration.
While empirical and theoretical evidence suggests that parameter-efficient and differentially private fine-tuning techniques can defend against privacy attacks on a fine-tuned model, PreCurious demonstrates the possibility of breaking up this invulnerability in a stealthy manner compared to fine-tuning on a benign pre-trained model.
While DP provides some mitigation for membership inference attack, by further leveraging a sanitized dataset, PreCurious demonstrates potential vulnerabilities for targeted data extraction even under differentially private tuning with a strict privacy budget e.g. $\epsilon=0.05$.
Thus, PreCurious raises warnings for users on the potential risks of  downloading pre-trained models from unknown sources, relying solely on tutorials or common-sense defenses, and releasing sanitized datasets even after perfect scrubbing.
\end{abstract}



\maketitle

\section{Introduction}\label{chap:intro}
The pre-training and fine-tuning paradigm has become the standard approach for tailoring language models (LMs) to various tasks, such as the medical domain~\cite{lee2020biobert, huang2019clinicalbert}.
In this approach, a language model is pre-trained on a large, general dataset and then fine-tuned on a smaller, domain-specific dataset.
Privacy risks arise when the fine-tuning data is private and the fine-tuned model can be accessed as a service~\cite{mireshghallah2022memorization}.
One realistic scenario is that a hospital fine-tunes a model using local Electronic Health Record (EHR) data and then shares the API with other hospitals that lack such expertise.
Existing works broadly explore the privacy risks of training data via black-box access of the model~\cite{shokri2017membership, carlini2022membership, carlini2021extracting}, which is also applicable to the fine-tuned model.

In this paper, we reveal a new privacy attack surface where an attacker aims to escalate the privacy risk of the fine-tuning data from a fine-tuned model by manipulating the pre-trained language model loaded by the user before fine-tuning and then getting the black-box access to the fine-tuned model.
This is realistic since anyone can publish models on community-based platforms (e.g., Huggingface~\cite{huggingface}, GitHub~\cite{github}) without stringent validation processes.
A fine-tuning user may inadvertently download an untrusted pre-trained model from compromised sources, especially when popular models have different variants on platforms like Hugging Face. For instance, a victim could make a typo during the download process or fall for a malicious higher-version package registered with the same name as a legitimate model.

\begin{figure}[t!]
\centering
    \includegraphics[trim=0 0 0 0, clip, width=0.47\textwidth]{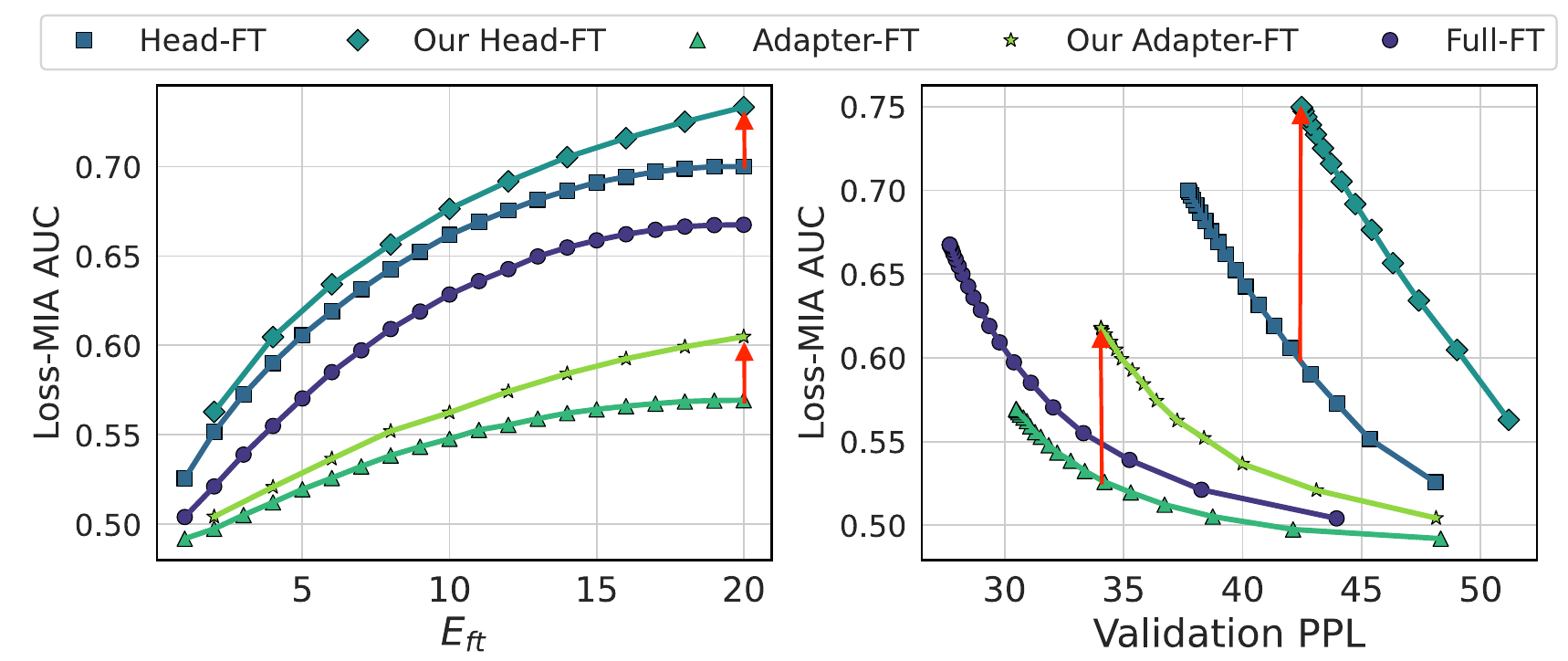}
	\caption{
    The privacy vulnerability for target models fine-tuned by various methods ranks as Head-FT $>$ Full $>$ Adapter-FT.
    \textit{PreCurious} increases the privacy risk for each iteration and ruins the privacy-utility trade-off, as demonstrated with Head-FT and Adapter-FT.
    $E_\text{ft}$ indicates the fine-tuning epochs and lower validation perplexity means better performance.
	}\label{fig:teaser}
\end{figure}

Previous work~\cite{tramer2022truth} explored additional adversarial access besides the black box access of the model by injecting poisoned data in the training dataset to amplify the privacy risk, which requires the adversarial capability of accessing/crafting training data.
A recent work~\cite{tian2023manipulating} manipulates pre-trained (upstream) image classification model for increasing the privacy risk of downstream models, but is limited to property inference attacks that infer whether images with a specific attribute are used for training.
In our threat model, the attacker aims to escalate the privacy risk by manipulating the released pre-trained model, without assuming access to the fine-tuning process or fine-tuning dataset.
Our adversarial goal is to amplify fundamental privacy threats of membership inference attack~\cite{carlini2022membership} and data extraction~\cite{carlini2021extracting} in the fine-tuned language model, compared to the one fine-tuned from a benign pre-trained model.

It is non-trivial to achieve our privacy risk amplification goal since parameter-efficient fine-tuning (PEFT) techniques such as Adapter~\cite{pfeiffer2020adapterhub} and LoRA~\cite{hu2021lora} have been established to have a privacy invulnerability property~\cite{wen2023last, mireshghallah2022memorization}. This is demonstrated in \Cref{fig:teaser} which shows the privacy vulnerability (measured in membership inference attack (MIA) effectiveness in AUC) for different fine-tuning methods vs. the fine-tuning epochs (left) and utility of the fine-tuned model (right) (measured in validation perplexity (PPL)). We can see that the Adaptor fine-tuning (Adaptor-FT) exhibit a very low vulnerability.
At the same time, the training efficiency introduced by PEFT~\cite{he2021towards, lialin2023scaling} makes it broadly applicable for LMs, especially encouraging differentially private (DP) fine-tuning for a large model~\cite{li2021large}, which makes the privacy attacks on the fine-tuned model more challenging.

\begin{figure*}[thb]
    \centering
    \includegraphics[width=0.9\textwidth]{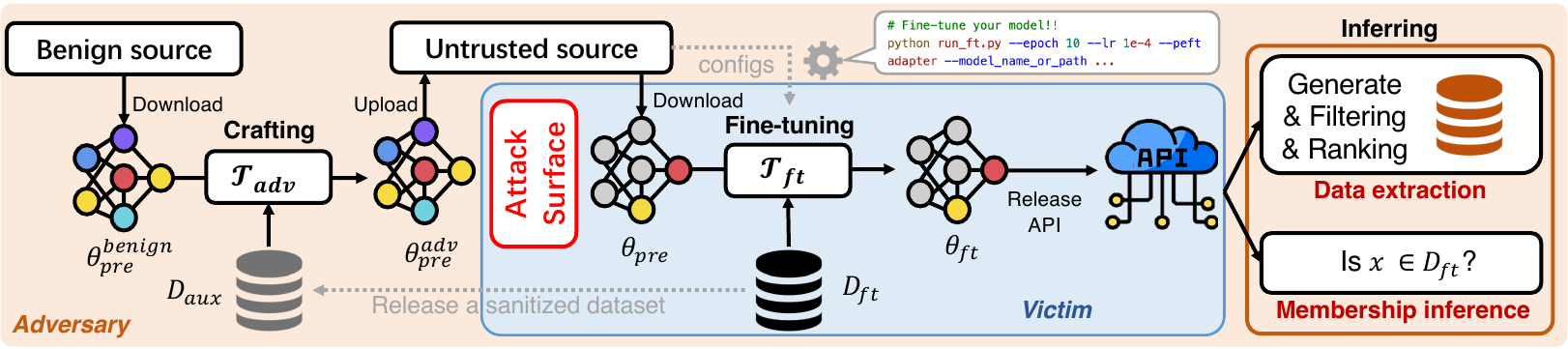}
    \caption{Framework overview of \textit{PreCurious}. 
    The dashed gray line indicates extra side information that can be utilized: 1) the stopping criterion, 2) the fine-tuning method, and 3) the released sanitized data by masking the secret.
    We design \textit{Accelerated} and \textit{Lagging} strategies for stopping by epoch or by performance.
    We propose an aggressive anti-freezing strategy when the victim uses the given fine-tuning method.
    We utilize a released sanitized dataset in targeted data extraction experiments.
    }
    \label{fig:framework}
\end{figure*}

Our key intuition is to manipulate the memorization level of the pre-trained model by exploiting PEFT.
Since the majority of the pre-trained model  is frozen during PEFT, we can better influence the behavior of the trainable modules for amplifying risks in the fine-tuned model. \Cref{fig:framework} illustrates our proposed framework where an attacker downloads a benign large model, manipulates it by an auxiliary dataset, and uploads it to an untrusted source for victims.
We exploit side information such as the stopping criterion and the fine-tuning method by implicitly guiding the fine-tuning victims through documents or tutorials, proposing \textit{lagging} or \textit{accelerating} strategies for cases with or without early stopping and \textit{anti-freezing} strategy when fine-tuning method is known.
Additionally, we attempt to make full use of the public information, for example, a released de-identified dataset, to further enhance the attack capability.

We demonstrate that our attack can successfully amplify various privacy risks. \Cref{fig:teaser} illustrates the increased privacy vulnerability of MIA by our methods on both Head-FT and Adaptor-FT. 
More generally, for MIA, we compare PreCurious with benign GPT-2~\cite{radford2019language} on the same black-box attack and demonstrate that by manipulating the pre-trained model, the true-positive-rate (TPR) at a false-positive-rate (FPR) of $0.01\%$  on Enron~\cite{data_enron}, PubMed~\cite{data_pubmed_cohan-etal-2018-discourse} and PTB~\cite{data-ptb-marcus-etal-1993-building} datasets is boosted by $8\times, 131\times$ and $36\times$, respectively.
For untargeted data extraction attack, we increase the times for a less duplicated sub-sequence shown in the pool of filtered generations by around $10\times$.
For targeted data extraction attack on Enron dataset, fine-tuning over benign model initialization cannot expose any secrets when fine-tuning with a strong DP level ($\epsilon=0.05$) while \textit{PreCurious} can extract 3 target email addresses with valid exposure values.
As advocated by previous work~\cite{tramer2022truth}, we also audit the stealthiness of \textit{PreCurious} and propose a mitigation method to make it more stealthy.

Our contribution can be summarized as follows:
\begin{itemize}
    \item We propose a framework \textit{PreCurious} to amplify the privacy risk of both membership inference and data extraction in the pre-training and fine-tuning paradigm, revealing the risk of fine-tuning over an unofficially released pre-trained LM.
    \item We propose two memorization manipulating strategies to craft the pre-trained model for fine-tuning with or without early-stopping. We further exploit the side-information of PEFT or sanitized dataset to enhance the attack effectiveness.
    \item We demonstrate the underestimated vulnerability of common-sense defenses, including regularization, differentially private fine-tuning, and deduplication with \textit{PreCurious}, particularly highlighting risks for users who rely on common-sense defenses without auditing privacy and training dynamics.
    \item We demonstrate the risks of publishing de-identified datasets solely by removing personally identifiable information (PII), as \textit{PreCurious} can exploit the context to extract targeted secrets if the original datasets are involved in future fine-tuning, underscoring significant vulnerabilities in the data release.
\end{itemize}

\section{Threat Model and Preliminaries}
We formulate the threat model and preliminaries in this section.
The attack framework of \textit{PreCurious} sits in the pre-training and fine-tuning paradigm of language models (LMs) to amplify data leakage in the fine-tuning stage.

Our target victim model is fine-tuned with the basic next-token prediction task.
The model aims to predict the next token $ x_{t}$ given the previous tokens $( x_1, x_2, ..., x_{t-1})$ for a given text sequence with $T$ tokens.
The fine-tuning involves minimizing the objective:
$\mathcal{L} = -\sum_{t=1}^{T} \log f_\theta(x_{t} | x_{i<t})$, where $f_\theta(x_{t} | x_{i<t})$ is the probability of $x_{t}$ from the softmax output of the model $\theta$.
The trained model can generate new text by iteratively sampling $\hat{x}_{t}\sim f_\theta(x_{t} | x_{i<t})$ and feeding it to sample the next token.

\subsection{Parameter-Efficient Fine-tuning (PEFT)}
Denoting the fine-tuned model as $\theta_\textup{ft}=\theta_\text{pre}\circ \Phi$, the key idea of PEFT is only optimizing over small modules $\Phi$ while freezing $\theta_\text{pre}$, which transfers the fine-tuning objective as $\mathcal{L} = -\sum_{t=1}^{T} \log f_\Phi(x_{t} | x_{i<t}, \theta_\text{pre})$.

One line of \textit{selective} PEFT  selects a portion of parameters in $\theta_\text{pre}$ as $\Phi$, such as Head-FT with a few top layers~\cite{donahue2014decaf} and Bitfit-FT with the bias terms of the model~\cite{zaken2021bitfit}.
The other line of PEFT introduces new randomly initialized modules as $\Phi$ as plug-in for $\theta_\text{pre}$.
For example, \textit{additive} method Adapter-FT~\cite{houlsby2019parameter} inserts small and trainable fully connected networks $\Phi$ after transformer sub-layers in $\theta_\text{pre}$.
The \textit{reparameterization-based} method LoRA-FT~\cite{hu2021lora} employs a low-rank matrix decomposition to parameterize the weight updates, and $\Phi$ indicates  parameters for the low-rank matrices.

\subsection{Threat Model}
\textit{PreCurious} indicates the \underline{pre}-trained model releaser is \underline{curious} about the private fine-tuning dataset $D_\text{ft} \in\mathcal{D}$.
We consider the model fine-tuner as the challenger $\mathcal{C}$ (or victim), and pre-trained model publisher as the adversary $\mathcal{A}$.

\subsubsection{Adversarial Capabilities}\label{sec:capability}
We make two common adversarial capability assumptions.
First, we follow a common assumption~\cite{ye2022enhanced, sablayrolles2019white, yeom2018privacy, mattern2023membership} that the adversary can query the loss value for a given sample via black-box access.
Second, following previous works~\cite{tramer2022truth, sablayrolles2019white, watson2021importance, mireshghallah2022memorization, shokri2017membership, jayaraman2020revisiting, ye2022enhanced}, we assume the adversary has an auxiliary dataset $D_\text{aux} \in \mathcal{D}$ drawn from the same distribution but disjoint from the fine-tuning dataset $D_\text{ft}$.
Different from capabilities in backdoor attacks on the pre-trained model, we do not assume either access to pre-training dataset of the original backbone~\cite{jia2022badencoder} or the access to the samples in downstream dataset~\cite{zhang2021trojaning}.
Additionally, we do not require capability of injecting poisoned data~\cite{tramer2022truth} or tampering the fine-tuning process.

Distinguished from all existing works, the adversary in \textit{PreCurious} releases the pre-trained model with seemingly legitimate configuration documents, which is very common when sharing customized models on open-sourced platforms.
We also note that even for popular pre-trained models, the victim may inadvertently download an untrusted $\theta_{pre}$.
In this case, attackers could use the official model's default configuration in tutorials, which victims assume as correct.
First, typographical errors during the search and download process, such as \lstinline[style=mystyle]{hf_hub_download(repo_id=NAME_WITH_TYPO)} in Hugging Face, could lead to the acquisition of a malicious model. 
Second, attackers could register publicly available higher-version packages with the same name as the legitimate model, which could be automatically installed via library management tools. 
Finally, the attacker could compromise the repository's infrastructure and replacing the legitimate pre-trained model with a malicious one.

The seemingly legitimate configuration $C=\{C_\text{stop},C_\text{peft}\}$ includes: 
1) stopping criterion $C_\text{stop}\in \{c_\text{epoch},c_\text{perf}\}$ of stopping-by-epochs or early-stopping-by-performance without imposing fixed hyper-parameters, and
2) PEFT strategy $C_\text{peft}$ like Adapter-FT or LoRA-FT that can be easily set using open-source frameworks~\cite{pfeiffer2020adapterhub}.
$C_\text{peft}$ is optional and only used for an accelerated variant in \Cref{sec:anti-freeze}.

We do not require the adversarial capability to pre-train a language model from scratch.
Thus, we assume the released pre-trained model $\theta_\text{pre}^\text{adv}$ is crafted from a benign model $\theta_\text{pre}^\text{benign}$ downloaded from a trusted source.

\subsubsection{Privacy Game}
Now we construct the general privacy game between a challenger $\mathcal{C}$ (the model fine-tuner) and an adversary $\mathcal{A}$ (the pre-trained model publisher) in Game \ref{def:privacy_game}.

\definecolor{lightgray}{gray}{0.9}
\begin{game}[Privacy game in \textit{PreCurious}]\label{def:privacy_game}
\indent
\begin{itemize}
    \item \textcolor{purple}{The adversary crafts and releases model with a suggested configuration $C$, $\theta_\text{pre}^\text{adv} \leftarrow \mathcal{T}_\text{adv}(D_\text{aux}| \theta_\text{pre}^\text{benign}, C)$.}
	\item The challenger samples a training dataset $D_\text{ft} \in \mathcal{D}$ and a secret $z\in \mathcal{U}$ (such that $D_\text{ft} \cap \mathcal{U} = \emptyset$), combining as $D_\text{ft} \leftarrow D_\text{ft} \cup \{z\}$
	\item The challenger \textcolor{purple}{loads $\theta_\text{pre}^\text{adv}$ as the model initialization, follows $C$} in fine-tuning and releases the black-box access to the final model $\theta_\text{ft}^\text{adv} \leftarrow \mathcal{T}_\text{ft}(D_\text{ft} |\textcolor{purple}{\theta_\text{pre}^\text{adv}, C})$.
	\item The adversary queries $\theta_\text{ft}^\text{adv}$ and emits a guess $\hat{z}\in \mathcal{U}$.
	\item The adversary wins the game if $\hat{z} = z$.
\end{itemize}
\end{game}
We use $\mathcal{U}$ to denote the secret universe of $D_\text{ft}$.
Removing the procedures in red and replacing $\theta_\text{pre}^\text{adv}$ with a benign model $\theta_\text{pre}^\text{benign}$ reduces Game \ref{def:privacy_game} to a conventional privacy game.

\subsubsection{Adversarial Goal}
The adversary aims to increase the privacy risk in the fine-tuning training dataset $D_\text{ft}$.
We focus on two representative privacy notions as follows:
\begin{itemize}[leftmargin=*]
	\item \textbf{Membership Priavcy}~\cite{shokri2017membership} is defined on the existence of a given sample in the fine-tuning dataset $D_\text{ft}$.
	\item \textbf{Extraction Privacy}~\cite{carlini2021extracting} is defined on the verbatim extraction of a subsequence in $D_\text{ft}$.
    The extraction is targeted if the attacker defines the format of secrets before the attack.
\end{itemize}

Concretely, $\mathcal{U}$ covers both membership privacy and extraction privacy by different instantiations.
For example, for membership inference, $\mathcal{U}=\{\mathbf{x}, \bot\}$ denotes two cases where a sample $\mathbf{x}$ is or is not in $D_\text{ft}$.
For data extraction, $\mathcal{U}$ consists of the collection of all candidate secrets for a piece of text in $D_\text{ft}$.

Furthermore, the adversary aims to amplify the privacy risk in $D_\text{ft}$ compared to fine-tuning from a benign model, as formally defined in Definition~\ref{def:success}.
\begin{definition}[Successful privacy risk amplification]\label{def:success}
    Given the same fine-tuning procedure $\mathcal{T}_\text{ft}$ based on a benign model $\theta_\text{pre}^\text{benign}$, and considering two privacy games differentiated by $\mathcal{T}_\text{adv}$ as $\mathcal{G} \simeq_{\mathcal{T}_\text{adv}} \mathcal{G}^\prime$,
    the privacy risk is amplified by $\mathcal{T}_\text{adv}$ when the adversarial gain:
    \begin{align}
        \Delta \text{Adv}_{\mathcal{G} \simeq_{\mathcal{T}_\text{adv}} \mathcal{G}^\prime}^\text{ft} &= 
        \text{Adv}_\mathcal{G}(\mathcal{A}, D_\text{ft}, \theta_\text{ft}^\text{adv}, z | \mathcal{T}_\text{adv}) \nonumber \\
        &- \text{Adv}_{\mathcal{G}^\prime}(\mathcal{A}, D_\text{ft}, \theta_\text{ft}^\text{benign}, z) > 0. \nonumber
    \end{align}

\end{definition}
The $\text{Adv}_\mathcal{G}(\mathcal{A}, \cdot)$ can be a success metric for reflecting the adversary's advantage for a specific attack, for example, $\text{Adv}_{\mathcal{G}_\text{MIA}}(\mathcal{A}, \cdot) = 2\cdot \text{Pr}[\hat{z}=z]-1$ for MIA~\cite{salem2023sok}.

Meanwhile, the adversary should avoid suspicions from victims that the pre-trained model $\theta_\text{pre}^\text{adv}$ will increase privacy risks in $D_\text{ft}$.
As defined in Definition~\ref{def:stealthiness}, we simulates the risk auditing based on the most ideal assumption for victims to have a benign model.
Note that Definition~\ref{def:success} is computed on the fine-tuned model, while Definition~\ref{def:stealthiness} is measured on the pre-trained model.

\begin{definition}[Privacy risk amplification stealthiness]\label{def:stealthiness}
    The pre-trained model $\theta_\text{pre}^\text{adv}$ output by a crafting algorithm $\mathcal{T}_\text{adv}$ is stealthy when the adversarial gain compared to $\theta_\text{pre}^\text{benign}$ satisfies:
    \begin{align}
    \Delta \text{Adv}_{\mathcal{G} \simeq_{\mathcal{T}_\text{adv}} \mathcal{G}^\prime}^\text{pre} &= 
        \text{Adv}_\mathcal{G}(\mathcal{A}, D_\text{ft}, \theta_\text{pre}^\text{adv}, z | \mathcal{T}_\text{adv}) \nonumber \\
        &- \text{Adv}_{\mathcal{G}^\prime}(\mathcal{A}, D_\text{ft}, \theta_\text{pre}^\text{benign}, z) \approx  0. 
        \nonumber
    \end{align}
\end{definition}

For stealth, the simplest but most effective way is not involving any fine-tuning samples in the crafting phase, which is consistent with the adversarial capabilities defined in \Cref{sec:capability} that $\mathcal{A}$ knows no exact samples in $D_\text{ft}$ and $D_\text{aux}$ is disjoint from $D_\text{ft}$.
As models cannot memorize secret before seeing it, the adversarial gain compared to the benign model for $D_\text{ft}$ should satisfy Definition~\ref{def:success}.

\subsection{Success Metrics}
Now we introduce concrete attack effectiveness metrics for different attacks and propose stealthiness metrics for victims to audit the pre-trained model.
\subsubsection{Membership Inference Attack}\label{sec:metric_mia}
We use AUC$~\uparrow$ to measure the effectiveness of the attack ( $\uparrow$ means the higher the value the more desirable the metric).
As suggested by previous work~\cite{carlini2022membership}, we also present results for MIA with TPR@FPR$\alpha$\%$~\uparrow$ given a small $\alpha$.
A lower $\alpha$ emphasizes the cost of false positives.

\subsubsection{Data Extraction Attack}
For untargeted data extraction, we follow previous work~\cite{lee2021deduplicating} to capture the portion $p_\text{ext}\uparrow$ of sub-sequences emitted by the target model that are included in the fine-tuning dataset $D_\text{ft}$.
For targeted data extraction, we use the  exposure~\cite{carlini2019secret} to measure if a targeted secret such as a phone number or email address can be reliably extracted.

\subsubsection{Stealthiness}\label{sec:metric}
Following Definition~\ref{def:stealthiness}, we propose three representative metrics as indicators of the adversarial gain, and the difference compared with a benign model reflects the stealthiness of the released model.

First, we simulate MIA with a non-membership dataset drawn from the same distribution to audit the stealthiness $S_\text{mia}$ by using MIA success metrics such as AUC in \Cref{sec:metric_mia}.

Second, for simulating the data extraction attack in an efficient way~\cite{carlini2022quantifying}, we use the $k$-extractable rate as $
S_\text{mem} = \frac{1}{n}\sum_i^n\mathbb{I}_{k\text{-extract}}(\mathbf{x}, \theta_\text{pre}),
$
where $\mathbb{I}_{k\text{-extract}}=1$ indicates if the model can generate the suffix $s$ given a $k$-length prefix $\mathbf{x}=[p||s]$.

Lastly, as overfitting is considered as an important cause of various privacy attacks, the victim may calculate the performance difference $S_\text{gap}$ between the training and validation dataset as a signal of overfitting:
$
S_\text{gap} =  \text{PPL}(D_\text{val}| \theta_\text{pre}) - \text{PPL}(D_\text{ft} | \theta_\text{pre}),
$
where PPL is a standard performance metric of LMs.

Assuming the benign model derives the baseline stealthiness metric $S_\text{mia} = 0.5$ for AUC, $S_\text{mem} = 0$, and  $S_\text{gap} \leq 0$, our goal is to ensure a low gap for  $\theta_\text{pre}^\text{adv}$ compared to the baseline.

\section{Amplifying Privacy Risk with PreCurious}
In this section, we introduce the \textit{PreCurious} framework shown in \Cref{fig:framework}, crafting methodologies, and the inference pipelines.

\subsection{Attack Overview}
\subsubsection{PreCurious Framework}
We begin with a high-level overview of the pre-curious attack which involves the following three stages.
1) \textbf{Crafting}: the adversary carefully crafts the backbone model before releasing it as a pre-trained model.
2) \textbf{Fine-tuning}: the victim initializes the model with the released parameters and starts normal fine-tuning over the private training dataset.
3) \textbf{Inferring}: the adversary queries the target model and guesses secrets in $D_\text{ft}$.

\textit{PreCurious} focuses on designing the crafting stage for increasing the attack advantage and thus stands as a general framework for a wide range of inferring strategies.

\begin{figure}
	\includegraphics[trim=0 220 0 0, clip, width=\columnwidth]{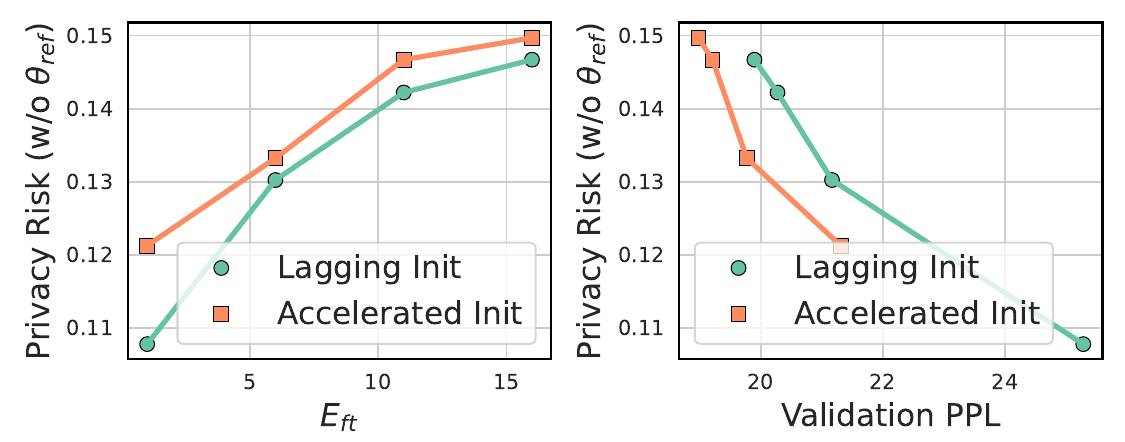}
	\caption{Privacy risk for different model initialization status.
	 Each point indicates the fine-tuned checkpoint for the Enron dataset with Adapter-FT.
	 We use TPR@0.1FPR as the proxy metric to measure the privacy risk of the model based on the scoring method in \Cref{eq:mia_loss}.
	We fully-finetuned the benign GPT-2 model on the auxiliary dataset for $E_\text{pre}=1$ and $E_\text{pre}=5$ separately for \textit{Lagging Init} and \textit{Accelerated Init} with learning rate $\eta_\text{pre}=10^{-5}$ as model initialization.
	}\label{fig:intuition}
\end{figure}

\subsubsection{Key Intuition}\label{sec:intuition}

From the feasible and limited capabilities in \Cref{sec:capability}, we notice that the one more thing that $\mathcal{A}$ can manipulate than a conventional attacker is the model initialization in the \textbf{crafting} stage.
Thus, we can first focus on the design of the model initialization in crafting and keep a basic inferring phase for now.

Based on previous lessons on memorization~\cite{carlini2022quantifying, mireshghallah2022memorization}, it is intuitive that using a better trained model as initialization induces overfitting on fine-tuning data, leading to higher privacy leakage via MIA or data extraction.
However, if we consider two models that achieve the same performance after fine-tuning, but one spends more iterations and the other spends fewer iterations, the intuition turns into the opposite: initializing with a less trained model may have a higher privacy risk because it will take more iterations for the model to achieve the same desired performance and the model will have seen the data more times and the influence of a sample is greater.
Both directions seem reasonable, we use the toy example in \Cref{fig:intuition} to show that the stopping criterion $\mathcal{C}$ is crucial for which intuition can lead to the success defined in \Cref{def:success}.

\textbf{Case I.} In our default setting with the criterion $c_\text{epoch}$, fine-tuning stops within arbitrary fixed epochs known only to the victim.
We expect a model initialization with higher memorization level leads to a higher privacy risk.
The \textbf{left} figure in \Cref{fig:intuition} confirms this intuition, since the privacy risk of \textit{Accelerated Init} given the same number of fine-tuning epochs is higher.

\textbf{Case II.} We consider another case where performance based early-stopping is used to avoid overfitting, for example, the fine-tuning stops when the validation performance achieves a certain level.
In the \textbf{right} figure of \Cref{fig:intuition},
we can observe that the \textit{Lagging Init} has a higher privacy risk given the same validation PPL.
Our insight is that a lagging initialization pushes fine-tuners to train more iterations for achieving the same performance, implicitly increasing the number of duplicates for training samples, which has been shown as a cause of higher privacy risk~\cite{lee2021deduplicating}.

By considering the stopping criterion when crafting the model initialization, our key intuition is to control the memorization stage for the model initialization on \textit{Lagging} and \textit{Accelerated} directions accordingly for achieving \Cref{def:success}.

\subsection{Methodology for Crafting}
Starting from the key intuition, we now introduce methodologies for controlling the two directions.
The accelerating by warm-up (\Cref{sec:basic}) and anti-freezing strategy (\Cref{sec:anti-freeze}) are proposed for \textbf{Case I} while the lagging strategy (\Cref{sec:lagging}) is proposed for \textbf{Case II}.

\subsubsection{Accelerating by Warm-up (Case I)}\label{sec:basic}
With no knowledge of specific PEFT methods in $\mathcal{T}_\text{ft}$, we propose a basic method  for accelerating the memorization stage in the fine-tuning data domain $\mathcal{D}$ by fully fine-tuning on $D_\text{aux}$.
Thus, for \textit{selective} PEFTs such as Head-FT or Bitfit-FT, the starting point for these trainable parts is already optimized for the domain $\mathcal{D}$, further tuning on these parameters can focus on learning the residuals or adjustments necessary to adapt the already domain-tuned representations of the base model to the nuances of $D_\text{ft}$.

For \textit{additive} PEFTs such as Adapter~\cite{houlsby2019parameter} and \textit{reparameterization-based} PEFTs such as LoRA~\cite{hu2021lora}, the inserted modules and low-rank matrices are usually randomly initialized by the victim.
It will take some iterations for these randomized parts to fit and enter the memorization-only stage, but it is still faster than fine-tuning on $\theta_\text{pre}^\text{benign}$ that is pre-trained over the out-of-domain public data.

\subsubsection{Accelerating by Anti-freezing (Case I)}\label{sec:anti-freeze}
When the victim follows the guidance provided by $\mathcal{A}$ on the choice of PEFT, $\mathcal{A}$ can utilize this side information for pushing the released model initialization $\theta_\text{pre}$ to the memorization-only stage with a more aggressive acceleration.

In typical \textit{addictive} and \textit{selective} PEFT training, only the small and random inserted modules are trainable while keeping the rest pre-trained parameters frozen.
On the contrary, we freeze the inserted / reparameterized modules and tune the backbone in our crafting stage.
The intuition is to make the released model equipped with a known PEFT module perfectly fit the data domain at the first step of $\mathcal{T}_\text{ft}$.
Thus, the first fine-tuning step enters the memorization-only stage~\cite{mireshghallah2022memorization} and the privacy risk will increase rapidly.

It should be noted that there is still a small amount of randomness because the PEFT modules initialized in $\mathcal{T}_\text{adv}$ by the adversary are different from the one initialized in $\mathcal{T}_\text{ft}$ by the victim if the random seed is not fixed.
Thus, we shift the seed in the two stages when performing the accelerated experiments for considering the influence of randomness.
By our observation, changing the seed causes subtle differences and does not affect the effectiveness, which may be because the randomly initialized modules are drawn from a common distribution.

\subsubsection{Lagging by Weight Scaling (Case II)}\label{sec:lagging}
In the opposite direction, for creating a lagging model initialization for privacy risk amplification in Case II, the intuitive idea is to make $\theta_\text{pre}$ perform worse or farther away from the data domain.

Ideally, learning a well-performed model is hard but hurting the utility is easy to achieve by simply spoiling the pre-trained parameters in $\theta_\text{pre}^{benign}$ with random noise, which does not even need the auxiliary knowledge $D_\text{aux}$.
However, an even perturbation on a well generalized pre-trained model cannot specifically manipulate the memorization stage on the fine-tuning domain.

For better control of the memorization stage towards the fine-tuning domain, we propose scaling a portion of parameters in the warmed-up backbone with a scaling factor $\beta$, which can be seen as an approximation of dropout~\cite{srivastava2014dropout}.
In each layer of a transformer-based backbone, there is a crucial component of multi-head self-attention (MHA).
Given a sequence of $l$ vectors $\mathbf{C}\in \mathbb{R}^{l\times d}$ and a query vector $\mathbf{q}\in \mathbb{R}^d$, the MHA output is:
\begin{align}
\text{Attn}(\mathbf{Q}, \mathbf{K}, \mathbf{V})&=\text{softmax}(\frac{\mathbf{Q}\mathbf{K}^\top}{\sqrt{d_k}})\mathbf{V}, \label{eq:attn} \\
\text{head}_i &= \text{Attn}(\mathbf{q}\mathbf{W}_q^{(i)}, \mathbf{C}\mathbf{W}_k^{(i)}, \mathbf{C}\mathbf{W}_v^{(i)}), \\
\text{MHA}(\mathbf{C}, \mathbf{q})&=\text{Concat}(\text{head}_1, \cdots, \text{head}_h)\mathbf{W}_o,
\end{align}
where $\mathbf{W}_q^{(i)}, \mathbf{W}_k^{(i)}, \mathbf{W}_v^{(i)} \in\mathbb{R}^{d\times d_h}$ and $\mathbf{W}_o \in \mathbb{R}^{d\times d}$.

Thus, if we use $\beta\in(0, 1)$ to scale weights $\mathbf{W}_q, \mathbf{W}_k, \mathbf{W}_v$, the magnitudes of the Q, K, and V vectors in \Cref{eq:attn} will decrease by a factor of $\beta$.
And the attention weights are more evenly distributed.
Additionally, scaling down $\mathbf{W}_o$  reduces the output magnitude and also hurts the expressiveness.
Therefore, a pre-trained model after weight scaling will result in a worse initial performance compared to a benign model.
We can apply the weight scaling strategy on the checkpoint after basic warm-up or after the accelerated strategy of anti-freezing for making the memorization degradation more specific to the domain $\mathcal{D}$.

On the one hand, it makes the model run more iterations to achieve the required performance.
On the other hand, the inserted small PEFT modules are encouraged to compensate for the reduced magnitude and expressiveness with a larger gradient magnitude.

\subsubsection{Rewinding for Stealthiness}
\label{sec:rewinding}
Since the victim might be suspicious of the crafting behavior, we propose to evade the abnormal values on proposed stealthiness metrics in \Cref{sec:metric}.
Rewinding~\cite{maini2023can} has been taken as a way to diagnose memorization in a neural network by replacing the weights of a single layer with an old version during training.

Our intuition for ensuring stealthiness is to find a controller for balancing the crafted version and a benign version.
Thus, for a crafted model $\theta_\text{pre}^{adv}$, we rewind a layer to its old version in $\theta_\text{pre}^{benign}$.
By controlling which layer and how many layers are rewound, we can trade off between stealthiness and attack effectiveness.

\subsection{Inference Pipeline}
\subsubsection{Membership Inference Pipeline}
In the inferring stage, we consider two standard membership scores for maximizing the adversary advantage in distinguishing the IN-world when $z=\mathbf{x}$ and OUT-world when $z=\bot$.

For the weakest adversary with no auxiliary dataset, loss value is a conventional signal for classifying samples as a member~\cite{yeom2018privacy}:
\begin{align}
A_\theta(\mathbf{x}) = \mathbb{I}[\mathcal{L}(\mathbf{x}; \theta) < \gamma].
\label{eq:mia_loss}
\end{align}
For an adversary with an auxiliary dataset or equally the predominant adversary $\mathcal{A}$ in our case, we follow the state-of-the-art attacks~\cite{sablayrolles2019white, carlini2022membership, ye2022enhanced, mireshghallah2022memorization} and calibrate the membership score with a difficulty score, which can be estimated with an OUT-world reference model $\theta_\text{ref}$ trained with the auxiliary dataset.
Thus, the signal for classification becomes:
\begin{align}
A_\theta(\mathbf{x}) = \mathbb{I}[\mathcal{L}(\mathbf{x}; \theta) - \mathcal{L}(\mathbf{x}; \theta_\text{ref}) < \gamma].
	\label{eq:mia_ref}
\end{align}

As previous works~\cite{mireshghallah2022memorization, mireshghallah2022quantifying}, we threshold the above two signals by setting $\gamma$ as the highest value of which the false positive rate over all samples would not exceed $\alpha$ for reporting the TPR with a given $\alpha$ FPR.
We omit the discussion on estimating the difficulty score by a pool of reference samples~\cite{mattern2023membership} because loss-value and reference-model scores have already covered the lower and upper bound of empirical MIA performance.
With the efficiency bottleneck on training multiple reference models, we limit the capability with only one reference model in all comparisons.

\subsubsection{Data Extraction Pipeline}\label{sec:method_extract}
We perform the data extraction in the inferring stage based on a state-of-the-art pipeline~\cite{carlini2021extracting} with two phases.
In the generation phase, the adversary will query the target model to generate a large amount of text with or without a given prefix.
In the membership inference phase, the adversary will sort the generated samples concerning \Cref{eq:mia_loss} or \Cref{eq:mia_ref} after deduplicating abnormally repeated samples.

\section{Experiments}
\subsection{Experimental Setup}
\textbf{Datasets.}
We run experiments on benchmark datasets from financial, email, and medical domains due to the confidential properties of the content, including Penn Treebank~\cite{data-ptb-marcus-etal-1993-building} (PTB), Enron~\cite{data_enron} and Pubmed~\cite{data_pubmed_cohan-etal-2018-discourse}.

We split the original training dataset equally into three partitions as $D_\text{ft}$, $D_\text{aux}$, and the non-member dataset $D_\text{non}$.
Thus, we avoid a false sense of attack effectiveness from the potential data shift~\cite{humphries2023investigating}.
To control the strength of this adversarial knowledge, we vary the data size ratio between the auxiliary dataset and the fine-tuning dataset $r_\text{aux} = |D_\text{aux}|/|D_\text{ft}|$ and by default $r_\text{aux}=1$ as the other work~\cite{tramer2022truth}.
For a fair comparison, we ensure same datasets are used in comparisons.

\textbf{Models and Parameter-Efficient Fine-Tuning.}
For the scalability to different backbone model sizes, we perform experiments on GPT-2 (12-layer, 117M), GPT-2-medium (24-layer, 345M), and GPT-2-large (36-layer, 774M) models.
Except for fully fine-tuning (Full-FT), we extend our evaluation to two \textit{selective} methods of Bitfit-FT and Head-FT, one \textit{addictive} method of Adapter-FT in the output layer with a reduction factor as $16$ and one \textit{reparameterization-based} method of LoRA-FT with $r=16$.
 
We set a default learning rate $\eta$ in Full-FT, Adapter-FT, LoRA-FT, Bitfit-FT, and Head-FT as $\{1e^{-5}, 1e^{-4}, 5e^{-4}, 5e^{-4}, 1e^{-4}\}$ with the linear scheduler in all baselines for a fair comparison.
By default, we train the model with $E_\text{ft}=20$ on GPT-2, $E_\text{ft}=5$ for GPT-2-medium/large and stop without overfitting.

\begin{table*}[thb]
\caption{Membership inference evaluation on GPT-2 with various PEFTs ($E_\text{ft}=20, E_\text{pre}=4$).
\texttt{Loss-Att} indicates loss-value based MIA in \Cref{eq:mia_loss} and \texttt{Full-Ref} indicates reference-model-based MIA in \Cref{eq:mia_ref}.
PreCurious shows amplified risk on all datasets, all PEFT methods in all MIA success metrics, while slightly increases the model performance measured by Val-PPL. 
PreCurious-Stealthy has an inferior attack performance than Basic but still amplifies risks compared to benign models.
}\label{tab:main_mia_ref}
\centering
\resizebox{\textwidth}{!}{%
\begin{tabular}{cc|c|ccc|c|ccc|c|ccc}
\toprule[1pt]
                 Dataset           &                                           & \multicolumn{4}{c}{Enron}                                      & \multicolumn{4}{c}{PubMed}                                     & \multicolumn{4}{c}{PTB}                                        \\
\midrule[1pt]
                   \textbf{Adapter-FT}         &                                           & Val-PPL                & AUC & @FPR1\% & @FPR0.01\% & Val-PPL                & AUC & @FPR1\% & @FPR0.01\% & Val-PPL                & AUC & @FPR1\% & @FPR0.01\% \\
\midrule[1pt]
\multicolumn{1}{c|}{\multirow{2}{*}{PreCurious}} 
& Basic                                     & \multicolumn{1}{c|}{17.19}                  & 92.89\% & 16.17\%     & 2.40\%        & 15.93                  & 99.59\% & 92.34\%     & 68.33\%        & 23.16                  & 99.79\% & 96.85\%     & 92.84\%        \\
                 \multicolumn{1}{c|}{}           & Stealthy                                  & \multicolumn{1}{c|}{17.86}                  & 82.42\% & 7.63\%      & 1.80\%        & 18.78                  & 60.74\% & 2.66\%      & 0.57\%        & 25.37                  & 93.00\% & 46.70\%     & 14.90\%        \\
\midrule[1pt]                        
\multicolumn{1}{c|}{\multirow{2}{*}{Benign}}     & Loss-Att                               & \multicolumn{1}{c|}{19.84} & 55.00\% & 1.05\%      & 0.00\%        & \multicolumn{1}{c|}{18.71} & 56.04\% & 1.47\%      & 0.00\%        & \multicolumn{1}{c|}{30.43} & 56.97\% & 2.58\%      & 2.29\%        \\
                      \multicolumn{1}{c|}{}      & Full-Ref                            & \multicolumn{1}{c|}{19.84}                      & 81.24\% & 8.53\%   & 0.30\%        &  \multicolumn{1}{c|}{18.71}                         & 75.25\% & 11.46\%   & 0.52\%        &    \multicolumn{1}{c|}{30.43}                       & 70.11\% & 16.62\%      & 2.58\%     \\
\midrule[1pt]
                   \textbf{Bitfit-FT}         &                                           & Val-PPL                & AUC & @FPR1\% & @FPR0.01\% & Val-PPL                & AUC & @FPR1\% & @FPR0.01\% & Val-PPL                & AUC & @FPR1\% & @FPR0.01\% \\
\midrule[1pt]
\multicolumn{1}{c|}{\multirow{2}{*}{PreCurious}} & Basic                                     & \multicolumn{1}{c|}{17.33}                  & 76.20\% & 3.89\%     & 0.75\%        & 16.00                  & 76.01\% & 6.70\%     & 1.62\%        & 23.18                  & 94.90\% & 50.72\%     & 40.40\%        \\
                 \multicolumn{1}{c|}{}           & Stealthy                                  & \multicolumn{1}{c|}{18.77}                  & 59.06\% & 3.89\%      & 0.45\%        & 17.00                  & 61.21\% & 3.80\%      & 0.19\%        & 25.99                  & 71.24\% & 5.16\%     & 1.72\%        \\
\midrule[1pt]                        
\multicolumn{1}{c|}{\multirow{2}{*}{Benign}}     & Loss-Att                               & \multicolumn{1}{c|}{22.07} & 52.55\% & 1.20\%      & 0.00\%        & 21.57 & 51.51\% & 1.19\%      & 0.00\%        & 35.74 & 52.14\% & 2.29\%      & 2.01\%        \\
                     \multicolumn{1}{c|}{}       & Full-Ref &  \multicolumn{1}{c|}{22.07}                      & 58.06\% & 4.64\%      & 0.15\%        &  \multicolumn{1}{c|}{21.57}                      & 55.08\% & 2.04\%      & 0.00\%        &   \multicolumn{1}{c|}{35.74}                     & 65.14\% & 6.02\%      & 0.86\%        \\
\midrule[1pt]
\textbf{LoRA-FT}         &                                           & Val-PPL                & AUC & @FPR1\% & @FPR0.01\% & Val-PPL                & AUC & @FPR1\% & @FPR0.01\% & Val-PPL                & AUC & @FPR1\% & @FPR0.01\% \\
\midrule[1pt]
\multicolumn{1}{c|}{\multirow{2}{*}{PreCurious}} & Basic                                     & \multicolumn{1}{c|}{17.06}                  & 93.76\% & 17.37\%     & 1.95\%        & 16.83                  & 94.12\% & 52.73\%     & 22.35\%        & 23.06                  & 99.94\% & 97.99\%     & 93.98\%        \\
                 \multicolumn{1}{c|}{}           & Stealthy                                  & \multicolumn{1}{c|}{17.97}                  & 81.38\% & 8.83\%      & 2.10\%        & 15.94                  & 99.72\% & 93.87\%      & 69.42\%        & 25.91                  & 91.48\% & 36.39\%     & 17.48\%        \\
\midrule[1pt]                        
\multicolumn{1}{c|}{\multirow{2}{*}{Benign}}     & Loss-Att                               & \multicolumn{1}{c|}{20.12} & 54.74\% & 1.05\%      & 0.00\%        & 19.24 & 55.86\% & 1.38\%      & 0.00\%        & 32.02 & 56.82\% & 2.87\%      & 2.29\%        \\
                     \multicolumn{1}{c|}{}       & Full-Ref &  \multicolumn{1}{c|}{20.12}                      & 75.96\% & 3.14\%      & 0.30\%        &  \multicolumn{1}{c|}{19.24}                      & 86.64\% & 26.63\%      & 0.38\%        &   \multicolumn{1}{c|}{32.02}                     & 85.30\% & 36.68\%      & 15.76\%        \\
\midrule[1pt]
                 \textbf{Head-FT}      &                                           & Val-PPL                & AUC & @FPR1\% & @FPR0.01\% & Val-PPL                & AUC & @FPR1\% & @FPR0.01\% & Val-PPL                & AUC & @FPR1\% & @FPR0.01\% \\
\midrule[1pt]
\multicolumn{1}{c|}{\multirow{2}{*}{PreCurious}} & Basic                                     & 18.56                  & 96.63\% & 21.71\%     & 2.40\%        & 17.69                  & 98.77\% & 80.93\%     & 24.49\%        & 28.06                  & 99.32\% & 74.79\%     & 47.85\%        \\
                \multicolumn{1}{c|}{}           & Stealthy                                  & 19.18                  & 94.41\% & 18.86\%      & 0.30\%        & 18.20                  & 95.35\% & 58.39\%      & 19.50\%        & 29.02                  & 99.70\% & 87.39\%     & 79.94\%        \\
\midrule[1pt]                        
\multicolumn{1}{c|}{\multirow{2}{*}{Benign}}     & Loss-Att                               & \multicolumn{1}{c|}{35.93}                  & 54.72\% & 1.20\%      & 0.00\%        & \multicolumn{1}{c|}{30.57}                  & 52.97\% & 1.24\%      & 0.00\%        & \multicolumn{1}{c|}{50.31}                  & 54.79\% & 3.44\%      & 1.72\%        \\
                     \multicolumn{1}{c|}{}      & Full-Ref                            & \multicolumn{1}{c|}{35.93}                                             & 57.26\% & 6.29\%   & 0.45\%        &  \multicolumn{1}{c|}{30.57}                         & 56.56\% & 0.02\%   & 0.00\%        &    \multicolumn{1}{c|}{50.31}                       & 68.18\% & 4.30\%      & 2.29\%     \\
\midrule[1pt]
                  \textbf{Full-FT}         &                                           & Val-PPL                & AUC & @FPR1\% & @FPR0.01\% & Val-PPL                & AUC & @FPR1\% & @FPR0.01\% & Val-PPL                & AUC & @FPR1\% & @FPR0.01\% \\
\midrule[1pt]
\multicolumn{1}{c|}{\multirow{2}{*}{PreCurious}} & Basic                                     & \multicolumn{1}{c|}{16.68}                  & 96.49\% & 30.24\%     & 1.95\%        & 15.46                  & 99.99\% & 100.00\%     & 99.95\%        & 22.31                  & 99.99\% & 100.00\%     & 99.43\%        \\
                \multicolumn{1}{c|}{}           & Stealthy                                  & \multicolumn{1}{c|}{16.84}                  & 96.17\% & 35.03\%      & 2.10\%        & 17.45                  & 72.92\% & 7.56\%      & 1.24\%        & 23.07                  & 99.97\% & 99.71\%     & 97.99\%        \\
\midrule[1pt]                        
\multicolumn{1}{c|}{\multirow{2}{*}{Benign}}     & Loss-Att                               & \multicolumn{1}{c|}{18.49} & 62.95\% & 1.20\%      & 0.00\%        & \multicolumn{1}{c|}{17.42} & 64.85\% & 1.81\%      & 0.00\%        & \multicolumn{1}{c|}{27.67} & 66.79\% & 4.58\%      & 2.87\%        \\
                     \multicolumn{1}{c|}{}      & Full-Ref                            & \multicolumn{1}{c|}{18.49}                      & 91.56\% & 14.22\%   & 1.35\%        &  \multicolumn{1}{c|}{17.42}                         & 98.93\% & 90.16\%   & 73.04\%        &    \multicolumn{1}{c|}{27.67}                       & 93.39\% & 66.48\%      & 64.18\%     \\
\bottomrule[1pt]  
\end{tabular}
}
\end{table*}

\begin{table*}[htb]
\caption{Membership inference evaluation on GPT-2 medium and GPT-2 large with AdapterFT ($E_\text{ft}=5, E_\text{pre}=3$)}
\label{tab:main_models}
\centering
\resizebox{\textwidth}{!}{%
\begin{tabular}{cc|c|ccc|c|ccc|c|ccc}
\toprule[1pt]
\multicolumn{2}{c}{Adapter-FT}     & \multicolumn{4}{c}{Enron}          & \multicolumn{4}{c}{PubMed}         & \multicolumn{4}{c}{PTB}             \\
\midrule[1pt]
\multicolumn{2}{c}{\textbf{GPT-2 Medium}} & Val-PPL & AUC & @FPR1\% & @FPR0.01\% & Val-PPL & AUC & @FPR1\% & @FPR0.01\% & Val-PPL & AUC & @FPR1\% & @FPR0.01\% \\
\midrule[1pt]
PreCurious              & Basic    & 14.18 & 84.31\% & 6.29\%  & 0.75\% & 13.01 & 96.48\% & 51.93\% & 2.38\% & 20.11 & 97.47\% & 67.05\% & 48.71\% \\
\midrule[1pt]
\multirow{2}{*}{Benign} & Loss-Att & 17.17 & 53.48\% & 1.20\%  & 0.15\% & 14.82 & 54.68\% & 1.19\%  & 0.00\% & 26.97 & 53.62\% & 1.72\%  & 1.15\%  \\
                        & Full-Ref & 17.17 & 58.12\% & 2.40\%  & 0.75\% & 14.82 & 73.39\% & 9.89\%  & 1.14\% & 26.97 & 62.81\% & 5.16\%  & 2.58\%  \\
\midrule[1pt]
\multicolumn{2}{c}{\textbf{GPT-2 Large}}  & Val-PPL & AUC & @FPR1\% & @FPR0.01\% & Val-PPL & AUC & @FPR1\% & @FPR0.01\% & Val-PPL & AUC & @FPR1\% & @FPR0.01\% \\
\midrule[1pt]
PreCurious              & Basic    & 12.39 & 87.24\% & 29.34\% & 5.54\% & 11.64 & 98.25\% & 73.99\% & 0.05\% & 16.94 & 99.40\% & 97.99\% & 96.56\% \\
\midrule[1pt]
\multirow{2}{*}{Benign} & Loss-Att & 14.92 & 57.01\% & 1.05\%  & 0.15\% & 12.82 & 59.47\% & 1.81\%  & 0.00\% & 21.66 & 60.79\% & 3.15\%  & 2.29\%  \\
                        & Full-Ref & 14.92 & 62.55\% & 6.44\%  & 2.25\% & 12.82 & 85.66\% & 24.68\% & 0.00\% & 21.66 & 78.78\% & 31.81\% & 24.07\% \\
\bottomrule[1pt]
\end{tabular}
}
\end{table*}

\begin{table*}[thb]
\caption{Membership inference evaluation on GPT-2 with Adapter-FT w/o $\theta_\text{ref}$ ($E_\text{ft}=20, E_\text{pre}=1$)}\label{tab:main_loss_mia_agg}
\centering
\resizebox{\textwidth}{!}{%
\begin{tabular}{c|c|ccc|c|ccc|c|ccc}
\toprule[1pt]
                 Dataset                                             & \multicolumn{4}{c}{Enron}                                      & \multicolumn{4}{c}{PubMed}                                     & \multicolumn{4}{c}{PTB}                                        \\
\midrule[1pt]
                   \textbf{Adapter-FT}  & Val-PPL                & AUC & @FPR1\% & @FPR0.01\% & Val-PPL                & AUC & @FPR1\% & @FPR0.01\% & Val-PPL                & AUC & @FPR1\% & @FPR0.01\% \\
\midrule[1pt]
\multicolumn{1}{c|}{\multirow{1}{*}{PreCurious-Accelerated}}                      & \multicolumn{1}{c|}{18.11}                  & 55.59\% & 1.20\%     & 0.00\%     
    & 16.08 & 56.78\% & 1.10\%     & 0.00\%        & 26.70                  & 58.03\% & 3.73\%     & 2.01\%        \\
\multicolumn{1}{c|}{\multirow{1}{*}{PreCurious-Basic}}                          & \multicolumn{1}{c|}{18.17}                  & 55.34\% & 1.20\%     & 0.00\%        & 16.09                  & 56.63\% & 1.19\%     & 0.00\%        & 26.54                  & 57.25\% & 3.15\%     & 1.72\%        \\
\multicolumn{1}{c|}{\multirow{1}{*}{Benign}}                           & \multicolumn{1}{c|}{19.84} & 55.00\% & 1.05\%      & 0.00\%        & \multicolumn{1}{c|}{18.71} & 56.04\% & 1.47\%      & 0.00\%        & \multicolumn{1}{c|}{30.43} & 56.97\% & 2.58\%      & 2.29\%        \\
\bottomrule[1pt]  
\end{tabular}
}
\end{table*}

\textbf{Baselines.}
For the main goal of verifying if  \textit{PreCurious} enlarges the adversarial gain as we defined in \Cref{def:success}, we compare the privacy risk of $\theta_\text{ft}^\text{adv}$ and $\theta_\text{ft}^\text{benign}$.

For all fine-tuned models, we use results w/ $\theta_\text{ref}$ to show risks for $\mathcal{A}$ who is the prominent adversary and the pretrained model publisher who has $D_\text{aux}$.
Results w/o $\theta_\text{ref}$ reflect risks from the potential weaker adversary $\mathcal{A}_w$ that can be anyone who queries the model but has no $D_\text{aux}$.
Thus, we could see the maximum secrets that can be inferred, as well as the attacking lower bound for the maximum coverage of potential adversaries.

As for $\theta_\text{ref}$, we use the model initialization as a default reference model, which is denoted as Base-Ref.
To control influence from calibration, we use $\theta_\text{ref}$ trained over the same $D_\text{aux}$ for benign baseline, which is denoted as Full-Ref.
By default, we evaluate baselines under \textbf{Case I} and discuss \textbf{Case II} in \Cref{sec:exp_tradeoff} for the early-stopping scenario.

\textbf{Metrics.}
We use the perplexity on validation dataset Val-PPL $\downarrow$ to measure the utility of the fine-tuned model.
As shown in \Cref{sec:metric}, we use  $S_\text{mia}\downarrow$,  $S_\text{mem}\downarrow$, and $S_\text{gap}\downarrow$  with suffix token length as $10$ to measure the stealthiness of the released model.
For privacy budget, we follow the widely applied setting $\delta=n^{-1.1}$ for all $\epsilon$\footnote{https://github.com/lxuechen/private-transformers.git}.
For AUC $\uparrow$ and TPR@FPR $\alpha$\%$\uparrow$ in MIA, we vary the FPR from 0.0001 to 0.1.
For untargeted data extraction, we vary the sub-sequence length by $L=\{2, 5, 10, 40, 50\}$.
For $v_\text{exp}\uparrow$ in targeted data extraction, we calculate the valid exposure threshold with the secret length of $L_\text{secret}=10$ characters.

\subsection{Effectiveness on Membership Inference}\label{sec:exp_mia}
In this section, we would like to measure the effectiveness of \textit{PreCurious} on amplifying the membership inference risk with the following questions:
\begin{itemize}[leftmargin=*]
	\item \textbf{RQ1:} What is the extent of the advantage gained through \textit{PreCurious} initialization compared to a benign one within the same iterations? (\Cref{sec:main_mia_rq1})
	\item \textbf{RQ2:} How does the choice of model initialization and reference model influence the adversarial advantage and interfere with each other? (\Cref{sec:grid_rq2})
	\item \textbf{RQ3:} Is the crafted backbone stealthy compared to the benign model? Which layer has more influence on stealthiness? (\Cref{sec:stealth_rq3})
	\item \textbf{RQ4}: Which conventional defenses fail on mitigating privacy risk when applying \textit{PreCurious}? (\Cref{sec:defense_rq4})
	\item \textbf{RQ5}: Does the risk amplification effect on MIA highly rely on the duplication between $D_\text{ft}$ and $D_\text{aux}$? (\Cref{sec:exp_dedup})
	\item \textbf{RQ6}: Can we break up the privacy-utility trade-off when early stopping is applied? (\Cref{sec:exp_tradeoff})
\end{itemize}

\begin{figure*}
	\includegraphics[width=\linewidth]{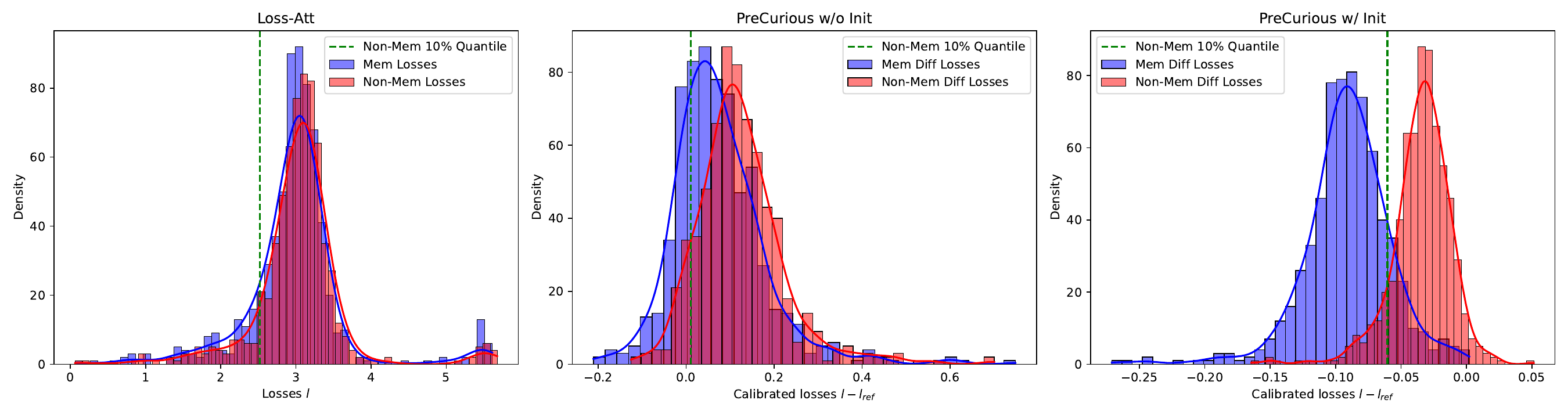}
    \vspace{-2em}
	\caption{Ablation study of PreCurious on the crafted initialization and reference model with Enron and Adapter-FT GPT-2.
    Loss distributions for Benign initialization w/o $\theta_\text{ref}$, benign initialization w/ Full-Ref, and PreCurious initialization w/ Full-Ref.
	}\label{fig:ablation}
\end{figure*}

\begin{figure}
	\includegraphics[trim=0 0 860 10, clip, width=0.75\linewidth]{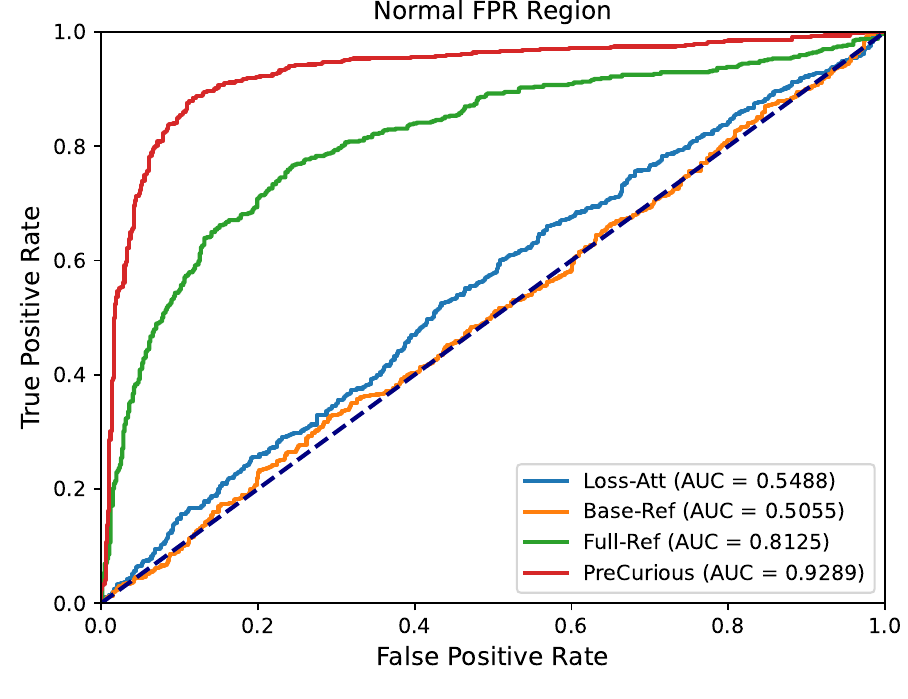}
    \vspace{-1.2em}
	\caption{ROC-AUC curve for Enron on Adapter-FT GPT-2.
    \texttt{Base-Full} indicates calibrating with a benign model cannot even beat \texttt{Loss-Att} with the same benign initialization.
	}\label{fig:ablation_auc}
\end{figure}

Denoting the learning rate, epochs in the \textbf{crafting} stage as $\eta_\text{pre}, E_\text{pre}$, we now clarify variants of \textit{PreCurious} as :
\begin{itemize}[leftmargin=*]
	\item \textbf{Basic} indicates the basic accelerating by warm-up (\Cref{sec:basic}).
	\item \textbf{Accelerated} indicates accelerating by anti-freezing  (\Cref{sec:anti-freeze}).
	\item \textbf{Lagging} means releasing the model with inferior performance on the domain (\Cref{sec:lagging}). By default, it means the combination of anti-freezing backbone and weight scaling.
 \item \textbf{Stealthy} is the stealthier version for \textit{Basic} by rewinding the head in the crafted backbone to the benign version (\Cref{sec:rewinding}).
\end{itemize}

\subsubsection{Performance Comparison}\label{sec:main_mia_rq1}
First, we summarize MIA performance between $\theta_\text{ft}^\text{benign}$ and $\theta_\text{ft}^\text{adv}$ in \Cref{tab:main_mia_ref} from the lens of the prominent adversary $\mathcal{A}$.
Using a $\theta_\text{ref}$ trained over $D_\text{aux}$ significantly improves the attacking effectiveness on the benign baseline as shown in previous works~\cite{tramer2022truth, mireshghallah2022memorization, sablayrolles2019white}.
Comparing with the state-of-the-art Full-Ref, we can see the adversary advantage is significantly amplified with a basic warm-up model initialization.
This is because the \textit{PreCurious}-Basic model initialization induces the fine-tuning process to start from a point close to the memorization-only stage~\cite{mireshghallah2022memorization} where membership inference risk increases rapidly and results in a higher privacy risk within given epochs.

Then, we evaluate the effectiveness of different backbones in \Cref{tab:main_models}.
We use the same reference model for Basic and Full-Ref for fair comparison, and we set $E_\text{ft}=5$ on the two larger models to avoid showing results after overfitting.
Comparing GPT-2 Medium with GPT-2 Large, under the same configurations, we can see that the Val-PPL and the MIA performance w/ or w/o $\theta_\text{ref}$ scales up with model size.
Comparing \textit{PreCurious}-Basic with Benign-Full-Ref, we can see that using a basic warm-up speeds up memorization and boosts the TPR@0.01\%FPR for PTB dataset by $\times 18.84$.

\begin{figure}[htb]
	\centering
	\begin{minipage}{0.25\textwidth}
        \centering
        \includegraphics[trim=12 5 25 25, clip, width=\linewidth]{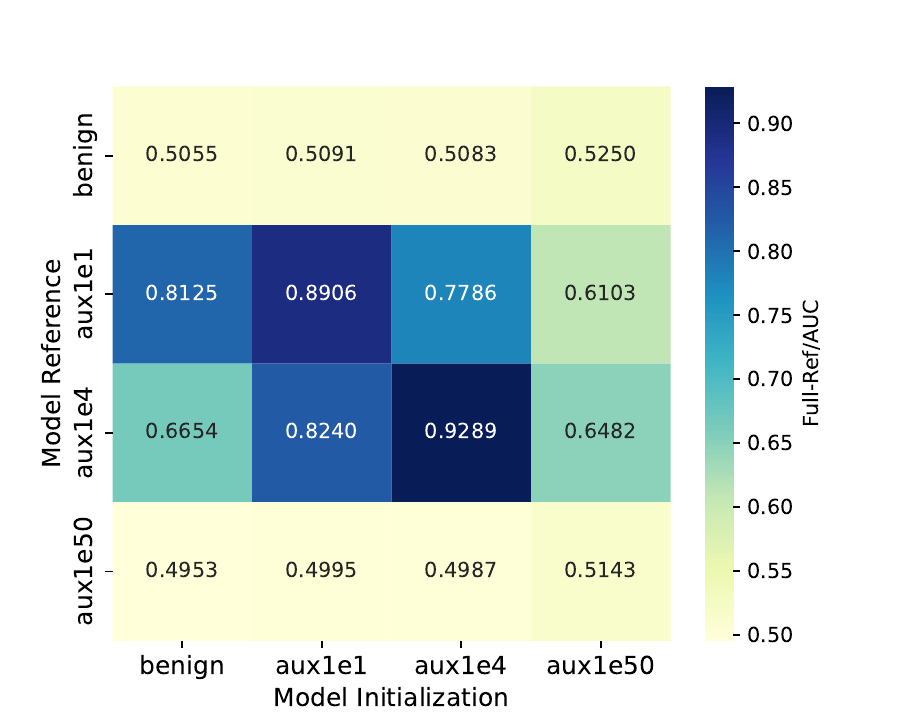}
    \end{minipage}%
    \begin{minipage}{0.25\textwidth}
        \centering
        \includegraphics[trim=12 5 25 25, clip, width=\linewidth]{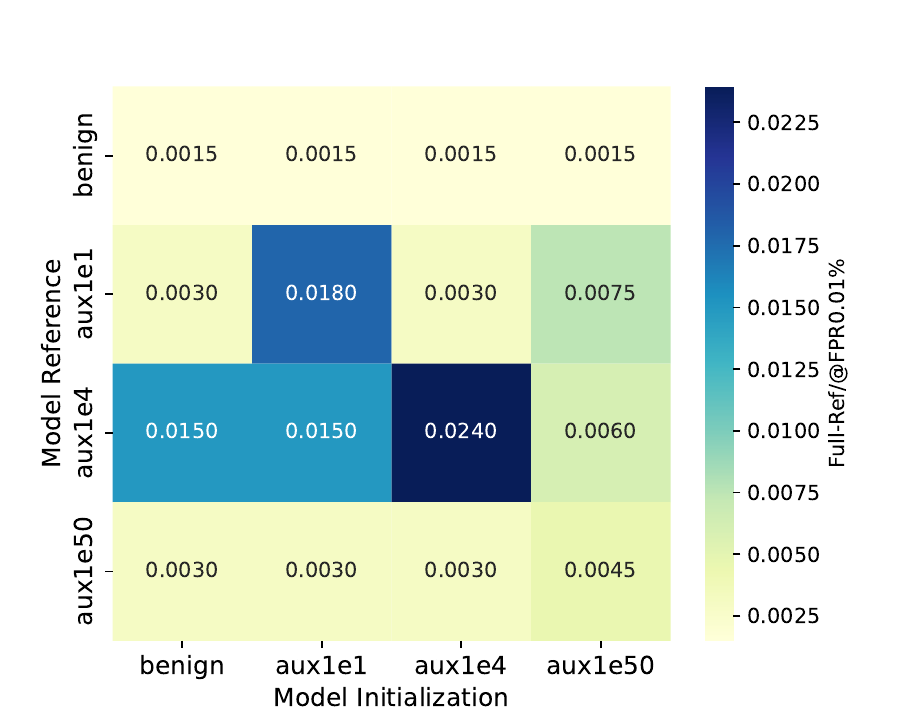}
    \end{minipage}%
    \caption{Influence of initialization and reference model choices on MIA success metrics (AdapterFT-Enron). \texttt{aux1e1} (under-fit), \texttt{aux1e4} (just-fit) and \texttt{aux1e50} (over-fit) denotes checkpoints warmed up on $D_\text{aux}$ with Full-FT in the crafting stage of \textit{PreCurious} to represent different overfitting levels on $D_\text{aux}$.
    We set a default  $\eta_\text{pre}=10^{-4}$ for fully fine-tuning in $\mathcal{T}_\text{pre}$ to reduce the required $E_\text{pre}$ when simulating the overfitting status here.
        }\label{fig:grid}
\end{figure} %

In addition, we observe the advantage introduced by model initialization in \Cref{tab:main_loss_mia_agg} by comparing Benign with Basic and the more aggressive Accelerated.
We set $E_\text{pre}=1$ as a safe choice for the accelerated version on all datasets.
There is a clear trend that the Val-PPL is decreasing and the privacy risk is increasing from Benign to Basic to Accelerated.
The Accelerated is indeed a more aggressive strategy that pushes the starting point to  memorization-only stage.

\textbf{RQ1-Response}: 
Whether with or without $\theta_\text{ref}$, the accelerated strategy of \textit{PreCurious} enhances the MIA advantage across different PEFTs and model sizes within the given number of iterations.

\subsubsection{Ablation Study}\label{sec:grid_rq2}
To show the independent advantage gained from the crafted initialization $\theta_\text{pre}^\text{adv}$ and the reference model $\theta_\text{ref}$, we perform an ablation study in \Cref{fig:ablation}, in which we choose the best reference model for achieving the highest MIA AUC on Benign-Full-Ref baseline.
First, the loss distribution shows the MIA signal distribution can be distinguished more significantly between members and non-members by adversarially crafting the initialization.
Then, comparing the ROC curve of PreCurious with Benign-Full-Ref, we can see the small advantage w/o $\theta_\text{ref}$ in \Cref{tab:main_loss_mia_agg} is amplified after calibration.
And we notice that the performance of calibration is highly sensitive to the choice of $\theta_\text{ref}$, as shown in \Cref{fig:ablation_auc}.
 
Now we would like to discuss the best choice of  $\theta_\text{pre}^\text{adv}$ and $\theta_\text{ref}$ for maximizing the MIA signal distinguishability, using PreCurious-Basic as an instance for the accelerated version.
To understand how different choice of model initialization and reference
model influence the adversarial advantage, we combine different warming-up checkpoints as $\theta_\text{ref}$ and $\theta_\text{pre}^\text{adv}$ in \Cref{fig:grid}.
First, we find a consistent rule that the best $\theta_\text{pre}^\text{adv}$ and $\theta_\text{ref}$ combination  for achieving the maximum advantage across different MIA metrics, datasets, and PEFTs is \texttt{aux1e4}-\texttt{aux1e4}.
Also, there is a clear trend that diagonal combinations yield higher risk, indicating the best $\theta_\text{ref}$ is $\theta_\text{pre}^\text{adv}$  or the one that has a slightly better performance to $\theta_\text{pre}^\text{adv}$.
Since the attack effectiveness of referenced model-based MIA is significantly influenced by the choice on $\theta_\text{ref}$, our finding solves the challenge by providing a simple rule of choosing $\theta_\text{ref}$.

\textbf{RQ2-Response}: 
$\mathcal{A}$ is suggested to use the just-fit model as $\theta_\text{ref}$ and $\theta_\text{pre}^\text{adv}$ in accelerated \textit{PreCurious}.

\begin{table}[thb]
\centering
\caption{Stealthiness on crafted $\theta_\text{pre}$. The red cell denotes `suspicious' and green cell indicates `evaded'.
}\label{fig:stealthy}
\resizebox{0.78\columnwidth}{!}{%
\begin{tabular}{c|c|ccc}
\toprule
Dataset & Released Model & $S_\text{mia}$ & $S_\text{mem}$ & $S_\text{gap}$ \\
\midrule
\multirow{5}{*}{Enron}  & Benign      & \cellcolor{gray!20}0.5130          & \cellcolor{gray!20}0.0359          & \cellcolor{gray!20}\textbf{-3.7130} \\
                        & Accelerated & \cellcolor{green!20}\textbf{0.5008} & \cellcolor{green!20}0.0255          & \cellcolor{red!20}{\ul -0.8853}    \\
                        & Basic       & \cellcolor{green!20}0.5054          & \cellcolor{red!20}{\ul 0.0494}    & \cellcolor{red!20}{\ul -0.8963}    \\
                        & Stealthy    & \cellcolor{green!20}0.5090          & \cellcolor{red!20}{\ul 0.0479}    & \cellcolor{red!20}{\ul -1.1640}    \\
                        & Lagging     & \cellcolor{green!20}\textbf{0.5008} & \cellcolor{green!20}\textbf{0.0000} & \cellcolor{red!20}{\ul 12.9240}    \\
\midrule
\multirow{5}{*}{Pubmed} & Benign      & \cellcolor{gray!20}\textbf{0.5010} & \cellcolor{gray!20}0.0005          & \cellcolor{gray!20}-0.0650          \\
                        & Accelerated & \cellcolor{red!20}{\ul 0.5084}    & \cellcolor{red!20}{\ul 0.0029}    & \cellcolor{green!20}-0.0940          \\
                        & Basic       & \cellcolor{red!20}{\ul 0.5071}    & \cellcolor{red!20}{\ul 0.0029}    & \cellcolor{green!20}-0.0974          \\
                        & Stealthy    & \cellcolor{red!20}{\ul 0.5060}    & \cellcolor{red!20}{\ul 0.0024}    & \cellcolor{green!20}-0.1105          \\
                        & Lagging     & \cellcolor{red!20}{\ul 0.5049}    & \cellcolor{green!20}\textbf{0.0000} & \cellcolor{green!20}\textbf{-1.2530} \\
\midrule
\multirow{5}{*}{Ptb}    & Benign      & \cellcolor{gray!20}0.4834          & \cellcolor{gray!20}0.0057          & \cellcolor{gray!20}6.5190           \\
                        & Accelerated & \cellcolor{green!20}\textbf{0.4805} & \cellcolor{red!20}{\ul 0.0086}    & \cellcolor{green!20}2.5140           \\
                        & Basic       & \cellcolor{green!20}0.4819          & \cellcolor{red!20}{\ul 0.0086}    & \cellcolor{green!20}3.0630  \\
                        & Stealthy    & \cellcolor{green!20}0.4816          & \cellcolor{red!20}{\ul 0.0086}    & \cellcolor{green!20}\textbf{2.3150}           \\
                        & Lagging     & \cellcolor{red!20}{\ul 0.5019}    & \cellcolor{green!20}\textbf{0.0000} & \cellcolor{green!20}3.8090          \\
\bottomrule
\end{tabular}
}
\end{table} %

\begin{figure}[htb]
	\includegraphics[width=\columnwidth]{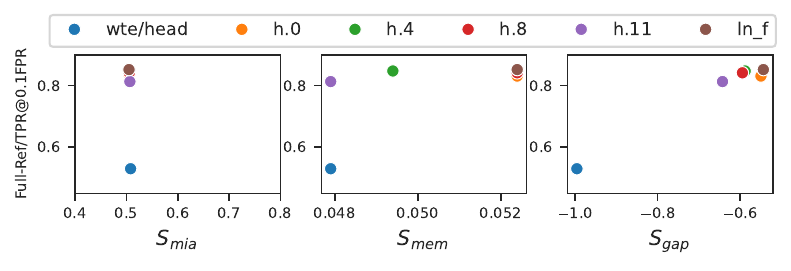}
	\vspace{-2.5em}
	\caption{Stealthiness-Risk trade-off via rewinding layers on Enron dataset with Adapter-FT.}\label{fig:stealthy_rewind}
\end{figure}

\subsubsection{Stealthiness}\label{sec:stealth_rq3}
Now we suppose the victim doubts the motivation of $\theta_\text{pre}^\text{adv}$ and the victim can query the benign $\theta_\text{benign}$ for auditing.
Thus, we compare the stealthiness metrics across benign backbone and \textit{PreCurious} backbones in \Cref{fig:stealthy}.
First, the proposed stealthiness metrics are possible to raise suspicion for $\theta_\text{pre}^\text{adv}$ if the victim is sensitive to the subtle differences.
$S_\text{mem}$ gives a more consistent detection compared to $S_\text{mia}$ or $S_\text{gap}$.
Second, \textit{Stealthy} is effective in enhancing the stealthiness of \textit{Basic}.
\textit{Accelerated} is also stealthier than the \textit{Basic} because auditing is performed on the backbone instead of composing with inserted modules.
But as shown in \Cref{tab:main_mia_ref}, \textit{Stealthy} sacrifices the attack effectiveness with the slight improvement on stealthiness.
Third, \textit{Lagging} has $S_\text{mem}=0$ and may successfully evade with  $S_\text{mia}\approx 0.5$ and low $S_\text{gap}$, except for $S_\text{gap}$ on Enron.
The high $S_\text{gap}$  results from the randomness of the poor initial utility.
Performing layer-wise rewinding in \Cref{fig:stealthy_rewind}, we observe that rewinding the last block can achieve the best stealthiness-risk trade-off.

\textbf{RQ3-Response}: 
\textit{PreCurious} increases stealthiness metrics very subtly and $\mathcal{A}$ can rewind the last block to further enhance the stealthiness.

\subsubsection{Effectiveness under Defense}\label{sec:defense_rq4}
\begin{figure}
\centering
	\includegraphics[trim=0 10 0 10, clip, width=0.25\textwidth]{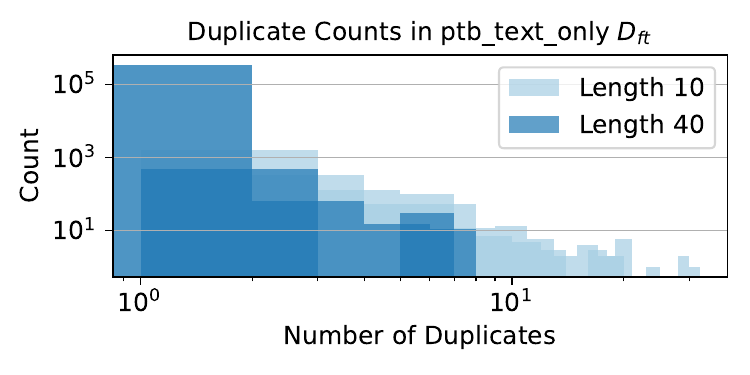}
	\includegraphics[trim=0 15 0 5, clip,width=0.3\textwidth]{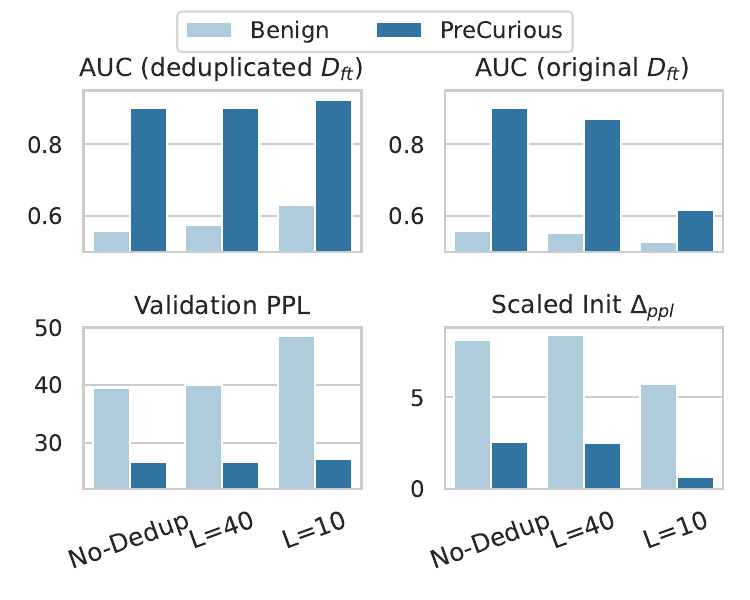}
	\caption{Duplication statistics and MIA effectiveness with Full-Ref under  deduplication defense on $D_\text{ft}$ for PTB.
		The stealthiness metric $S_\text{gap}=\Delta_\text{ppl}$ is linearly scaled for clear visualization.
		We randomly subsample non-membership samples for keeping the same size as the deduplicated $D_\text{ft}$ in MIA evaluation.
		}
	\label{fig:defense_deduplication}
\end{figure}

\begin{table}[ht]
\centering
\caption{MIA effectiveness under weight-decaying on Enron dataset with LoRA-FT (w/ weight decay factor 0.5).
}\label{tab:defense_wd_enron_lora}
\vspace{-0.5em}
\resizebox{\columnwidth}{!}{%
\begin{tabular}{c|cccc|c|c}
\toprule[1pt]
Model Init. & AUC w/o $\theta_\text{ref}$ & @0.01FPR & @0.1FPR & AUC     & Tr-PPL & Val-PPL \\
\midrule[1pt]
Benign      & 54.37\%                     & 2.40\%  & 38.32\%          & 73.48\% & 20.18           & 20.19  \\
PreCurious  & \textbf{55.18}\%          & \textbf{15.57\%}            & \textbf{85.63\%}           & \textbf{92.70\%} & \textbf{16.61}  & \textbf{17.07}   \\
\bottomrule[1pt]
\end{tabular}
}
\end{table}

\begin{table}[ht]
\centering
\caption{MIA effectiveness under DP fine-tuning defense on PTB dataset with Adapter-FT ($\epsilon=1$).
}
\label{tab:defense_dp_ptb_adapter}
\resizebox{0.8\columnwidth}{!}{%
\begin{tabular}{c|c|ccc|c}
\toprule[1pt]
Model Init. & Strategy & @0.01FPR & @0.1FPR & AUC     & Val-PPL    \\
\midrule[1pt]
Benign    & Full-Ref     & \textbf{1.72\%} & 10.03\%          & 52.05\%          & 68.61          \\
PreCurious    & Basic & 0.86\%          & \textbf{14.04}\%          & \textbf{54.84\%} & \textbf{25.94} \\
\bottomrule[1pt]
\end{tabular}
}
\end{table}

\begin{figure}[thb]
	\centering
	\includegraphics[trim=0 0 0 0, clip, width=0.38\textwidth]{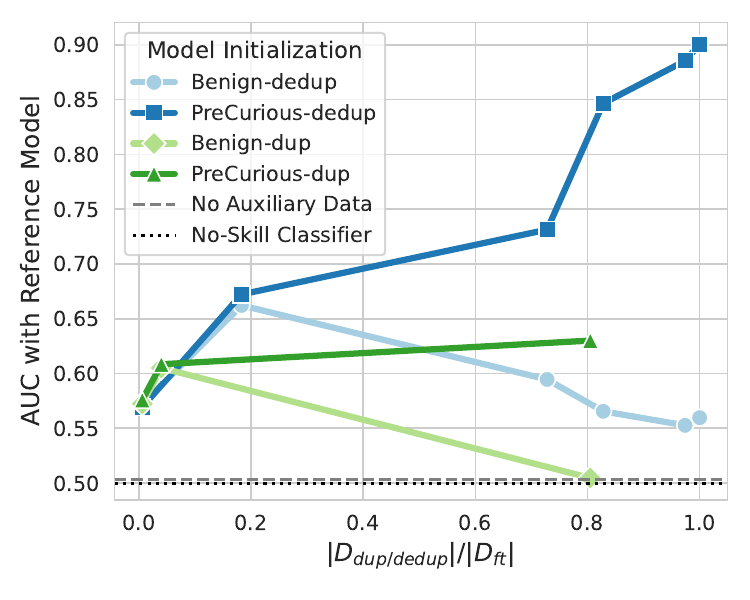}
	\vspace{-1.2em} %
	\caption{MIA effectiveness on PTB dataset with $D_\text{aux}^\text{dedup}$ and $D_\text{aux}^\text{dup}$ as auxiliary data for training $\theta_\text{ref}$ in Benign or $\theta_\text{pre/ref}$ in PreCurious, and  $|D_\text{dedup/dup}|/|D_\text{ft}|=1$ denotes the default $D_\text{aux}$ w/o deduplication. 
	}
	\label{fig:aux_deduplication}
\end{figure}

Under the representative defense strategy of weight decay, we show in \Cref{tab:defense_wd_enron_lora} that \textit{PreCurious} is robust on privacy risk amplification even with a high coefficient that exceeds the typical selection.

Under the strict defense of DP fine-tuning~\cite{yu2021differentially, li2021large}, 
we show in \Cref{tab:defense_dp_ptb_adapter} that \textit{PreCurious} model increases the AUC compared to the Benign model but has a smaller TPR@0.01FPR and better utility due to the warming-up.
The overall risk compared to non-DP fine-tuning in \Cref{tab:main_mia_ref} is significantly mitigated by DP, supported by more results w.r.t. various budgets in the Appendix \Cref{tab:defense_dp_ptb_adapter_APP}.

In \Cref{fig:defense_deduplication}, we evaluate the MIA effectivenss of Benign and PreCurious under deduplication defense~\cite{lee2021deduplicating, kandpal2022deduplicating}.
As shown in the duplicate statistics at the top, a sub-sequence in $D_\text{ft}$ may appear multiple times and make it easier to memorize~\cite {kandpal2022deduplicating}.
Deduplication can be instantiated with suffix array-based algorithm~\cite{lee2021deduplicating} for finding and mitigating repeated sub-sequences in $D_\text{ft}$.

By deduplicating repeated sub-sequence of length $L=\{10, 40\}$ in $D_\text{ft}$, we find a consistent trend that \textit{PreCurious} still causes a higher MIA risk than Benign initialization.
Taking original $D_\text{ft}$ as members, heavier deduplication leads to less privacy risk.
But we note that \textit{PreCurious} with a heavy deduplication such as $L=10$ still causes more privacy leakage than Benign baseline without deduplication.
Also, deduplication  helps $\mathcal{A}$ to be more stealthier and results in a higher perplexity (worse utility-privacy trade-off), because the auxiliary dataset is not deduplicated.
When taking samples in deduplicated $D_\text{ft}$ as members, the MIA risk is increasing for a heavier deduplication due to a larger distribution shift.
This is also because the data size used for fine-tuning is diminished and the deduplication essentially induces training samples to become outliers and more vulnerable to be inferred~\cite{tramer2022truth}.
The ideal case where attackers can approximate deduplicated texts in MIA inference can be seen as a corner case for deduplication defense to fail.

\textbf{RQ4-Response}: 
\textit{PreCurious} still effectively amplifies the privacy risk under defenses and is even stealthier under deduplication.

\subsubsection{Duplicates Investigation}\label{sec:exp_dedup}
In previous experiments, we use a randomly split dataset as $D_\text{aux}$ for launching \textit{PreCurious}.
However, $D_\text{aux}$ may have partially overlapped sub-sequence as in $D_\text{ft}$, which might be the reason for a successful privacy risk amplification. %
To understand whether the risk amplification effect is highly dependent on
the duplication between the two datasets $D_\text{ft}$ and $D_\text{aux}$, we control the overlapping level of $D_\text{aux}$ with cross-deduplication:
\begin{itemize}[leftmargin=*]
	\item For $D_\text{aux}^\text{dedup}$, we \textbf{drop} all $L$-length sub-sequences that overlaps with $D_\text{ft}$ on the default $D_\text{aux}$.
	\item For $D_\text{aux}^\text{dup}$, we find all cross-duplicated $L$-length sub-sequences and \textbf{keep} them to construct it.
\end{itemize}
By varying over different $L=\{2,5,10,40,60\}$, we get $D_\text{aux}^\text{dup}$ and $D_\text{aux}^\text{dedup}$ with various auxiliary dataset sizes.
It should be noted that this experiment is designed for analysis instead of a ``real'' attack as we are manipulating the adversary capability with $D_\text{ft}$.

\begin{figure}[thb]
\centering
\vspace{-0.8em}
\includegraphics[width=0.49\columnwidth]{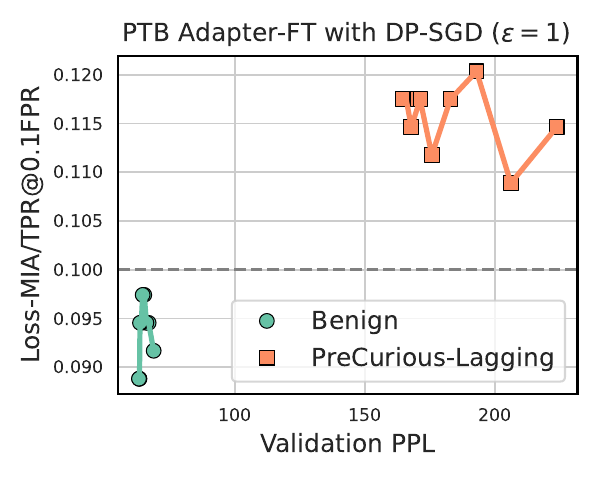}
\includegraphics[width=0.49\columnwidth]{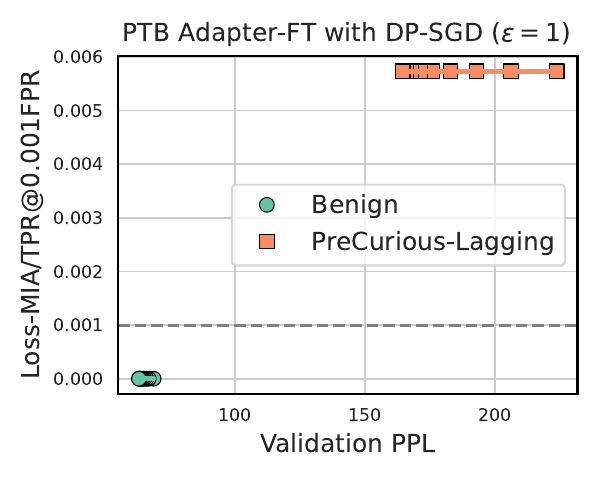}
\vspace{-1.2em}
\caption{Breaking up privacy-utility trade-off under DP.}\label{fig:defense_dp_tradeoff}
\end{figure}

\begin{figure}[thb]
\centering
\vspace{-0.8em}
	\includegraphics[width=\columnwidth]{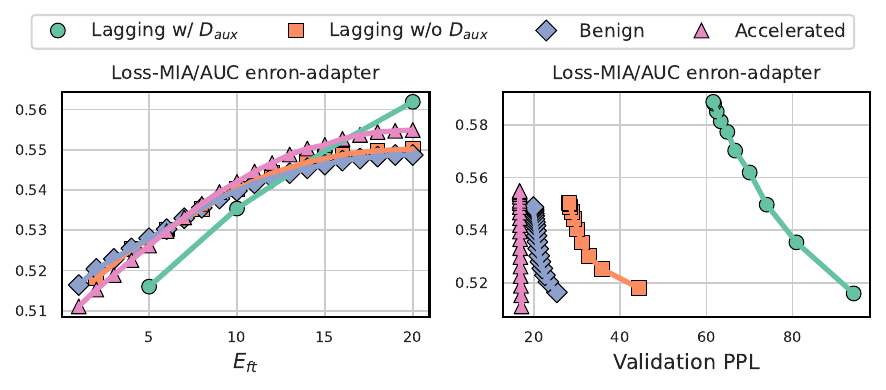}
\vspace{-1.8em}
\caption{MIA effectiveness for Enron and PTB datasets with Adapter-FT.
The baseline of Lagging w/ $D_\text{aux}$ indicates anti-freezing on $D_\text{aux}$ and then applying weight scaling with $\beta=0.1$.
We use different seeds when randomly initializing adapter module parameters for $\mathcal{T}_\text{pre}$ and $\mathcal{T}$.
Lagging w/o $D_\text{aux}$ performs the weight scaling directly on the benign  $\theta_\text{benign}$.
}\label{fig:tradeoff}
\end{figure}

As shown in \Cref{fig:aux_deduplication}, we control the duplication level by increasing $L$ for $D_\text{aux}^\text{dedup}$ and decreasing $L$ for $D_\text{aux}^\text{dup}$ from left to right.
We can observe that using the auxiliary knowledge with $D_\text{aux}^\text{dedup}$ has superior attack performance than $D_\text{aux}^\text{dup}$, which indicates that the privacy risk amplification of \textit{PreCurious} does not solely rely on the cross-duplicated parts between $D_\text{aux}$ and $D_\text{ft}$.
Then, we observe a clear trend for all datasets that the adversarial advantage of \textit{PreCurious} with auxiliary knowledge $D_\text{aux}^\text{dedup}$ increases with a moderate level of cross-deduplication, with a similar trend shown for Benign baseline with $\theta_\text{ref}$.
In addition, by only using the duplicated parts, which are typically the very common sub-sequences in the domain $\mathcal{D}$,
even the adversarial gain from $\theta_\text{ref}$ is poor, warming up with a batch of common fragments also helps to amplify the MIA risk, which weakens the required assumption on $D_\text{aux}$.

\textbf{RQ5-Response}: 
\textit{PreCurious} does not heavily rely on the duplicates between $D_\text{ft}$ and $D_\text{aux}$.

\begin{figure}[thb]
\centering
\vspace{-0.3em}
    \includegraphics[trim=10 0 0 0, clip, width=0.48\textwidth]{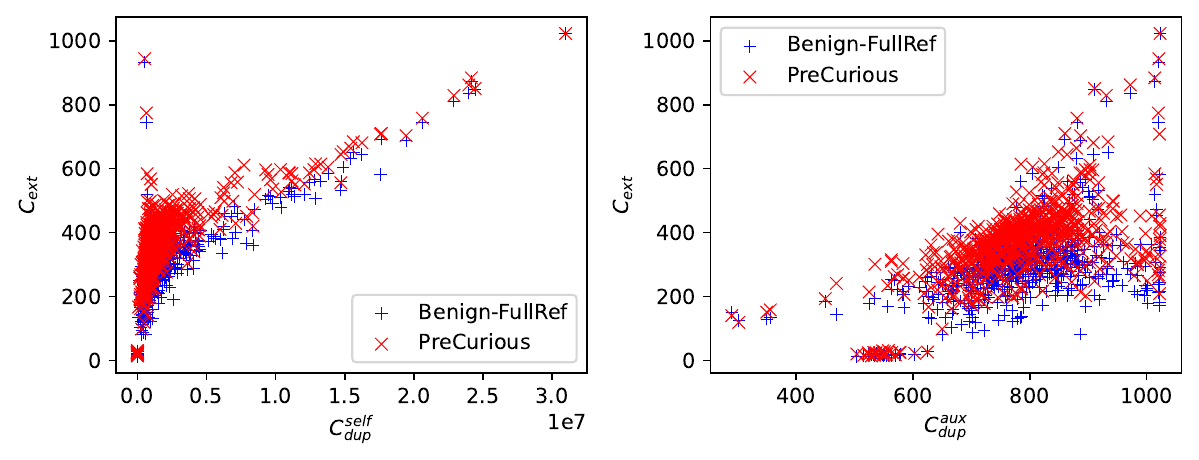}
    \includegraphics[trim=10 0 0 0, clip, width=0.48\textwidth]{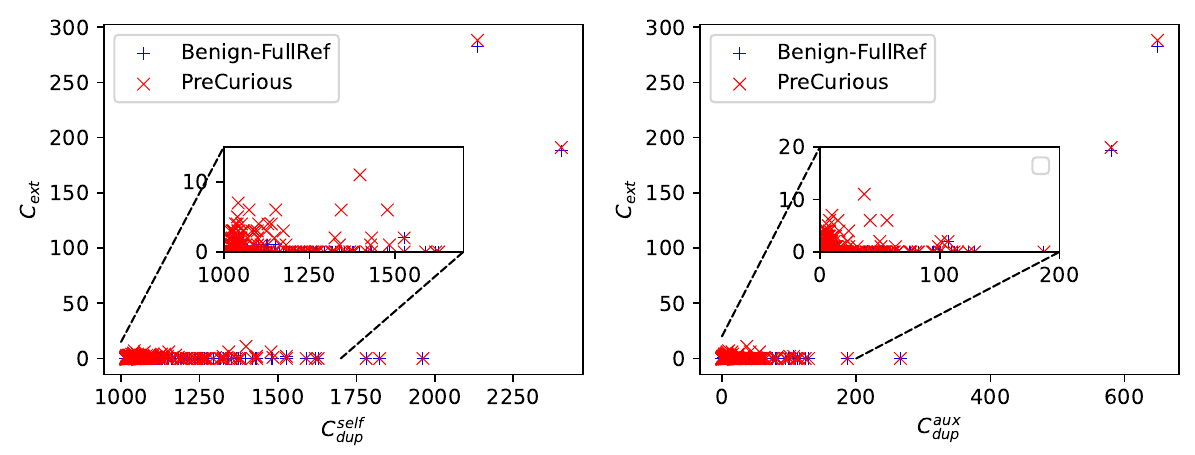}
\vspace{-1.8em}
 \caption{Untargeted Data Extraction for Adapter-FT model with $L_\text{sub}=2$ for for Enron (top) and $L_\text{sub}=10$ PTB (bottom).}\label{fig:untargeted}
\end{figure}

\begin{figure}[thb]
	\centering
 \includegraphics[trim=0 12 0 25, clip, width=0.9\columnwidth]{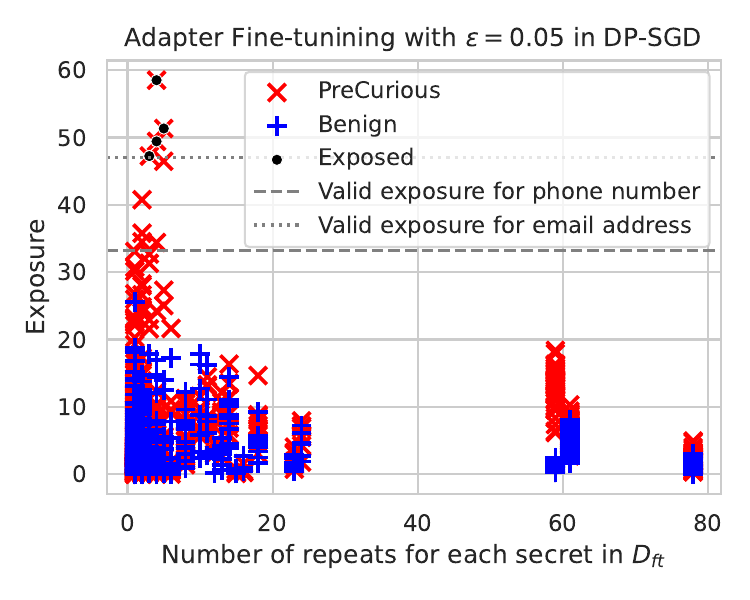}
	\caption{Targeted data extraction on Enron with Adapter-FT and $\epsilon=0.05$ for DP-SGD. No secret's exposure is above the valid threshold for fine-tuned benign model under DP.}\label{fig:targeted}
\end{figure}

\subsubsection{Breaking-up the trade-off}\label{sec:exp_tradeoff}
As shown in \Cref{fig:tradeoff}, we can use lagging \textit{PreCurious} to break up the privacy-utility trade-off and amplify the risk for Case II.
We compare all baselines with loss signals to avoid the influence of $\theta_\text{ref}$.
We can observe that \textit{PreCurious}-Lagging w/ $D_\text{aux}$ is possible to amplify the risk.
But only weight scaling on a benign backbone is not as effective as scaling with the same level on a warmed-up model to distinguish the loss signal distribution at the end, validating the effectiveness of anti-freezing.

It is seen that \textit{PreCurious}-Accelerated shows a consistent tendency to amplify risk given fixed epochs $E_\text{ft}$.
While \textit{PreCurious}-Lagging is robust in breaking up the privacy-utility trade-off, resulting in either poor model performance or high privacy risk, which validates our key intuition of increasing risk by increasing the required iterations to achieve the same utility.
One different observation is that applying a lagging initialization for LoRA-FT does not show the same sign to amplify risk given a fixed epoch as expected.
In addition, we find weight scaling with $\beta=0.1$ on \texttt{attn.c\_attn.weight} is effective while the effective choice for Adapter-FT is \texttt{attn.c\_proj.weight}, which are exactly where PEFT modules are applied, indicating the importance of fine-tuning side-information for the lagging strategy.

In addition, we address the privacy-utility trade-off issue in \Cref{tab:defense_dp_ptb_adapter} with the lagging strategy as shown in \Cref{fig:defense_dp_tradeoff}.
Even when the worst-case privacy is bounded by a strict DP, we show that $\epsilon=1$ is still not a perfect protection.
This success is due to more iterations for achieving the same utility, and also because 
the larger gradient norm derived from \textit{PreCurious}-lagging fully exploits the per-sample sensitivity to reflect the influence of each sample.

\textbf{RQ6-Response}: 
$\mathcal{A}$ is suggested to apply Lagging-\textit{PreCurious} for breaking-up utility-privacy trade-off when early stopping is applied.

\subsection{Effectiveness on Data Extraction}
Now we evaluate the effectiveness of \textit{PreCurious} on data extraction.
As previous work~\cite{kandpal2022deduplicating, lee2021deduplicating, carlini2019secret} conclude, less duplicated secrets are more challenging to be extracted, thus we raise questions:
\begin{itemize}[leftmargin=*]
	\item \textbf{RQ7}: Are less deduplicated training samples safe with DP training and constraint of limited query times? (\Cref{sec:untargeted})
	\item \textbf{RQ8}: How bad is \textit{PreCuious} when maximizing the auxiliary knowledge? (\Cref{sec:targeted})
\end{itemize}

\subsubsection{Untargeted Extraction}\label{sec:untargeted}
For \textbf{RQ7}, we focus on the effectiveness of samples of less duplication in $D_\text{ft}$ and assume the victim applies DP fine-tuning with $\epsilon=0.05$ and the target can only query for limited $1,000$ generations.
We perform the untargeted extraction in \Cref{sec:method_extract} for both Benign and \textit{PreCurious} by: 1) generating samples with a maximum length of $512$ via length $200$-length prefixes, and 2)  deduplicating and ranking by MIA signals in \Cref{eq:mia_ref} to filter $100$ samples.
The prefixes are constructed by using the top frequent phrases shown in $D_\text{aux}$ as we suppose the short but common parts can be transferred to $D_\text{ft}$.

In \Cref{fig:untargeted}, we use the $C_\text{ext}$ to denote the extraction level for \textbf{each sample} in $D_\text{ft}$, which counts the total times of its sub-sequences shown in all generated outputs.
The averaged performance measured by $p_\text{ext}$ is shown in Appendix \Cref{tab:un_ptb_adapter_APP}.
$C_\text{dup}^\text{self}$ and $C_\text{dup}^\text{aux}$ indicate the total times of its sub-sequences shown in $D_\text{ft}$ and $D_\text{aux}$, respectively.
In \Cref{fig:untargeted}, there is a clear trend that $C_\text{ext}$ increases with larger $D_\text{dup}^\text{self}$ and $D_\text{dup}^\text{aux}$, thus extracting less duplicates are indeed more challenging.
But \textit{PreCurious} can significantly improve the success on less duplicated samples, even under strict privacy defense given limited query times.

\textbf{RQ7-Response}: 
\textit{PreCurious} can still increase leakage of fewer-duplicated secrets even with DP fine-tuning.

\subsubsection{Targeted Extraction}\label{sec:targeted}
To investigate the threat when $\mathcal{A}$ in \textit{PreCurious}, we design the targeted extraction with the Enron dataset and take the phone number and email addresses as our targeted secrets.
For maximizing the auxiliary knowledge, we take a masked version of $D_\text{ft}$ as the $D_\text{aux}$, which is bold but possible because releasing de-identified text data is taken as a common practice~\cite{johnson2023mimic}.
After that, we apply \textit{PreCurious}-Basic and evaluate the exposure on our targeted secretes for both $\theta_\text{ft}^\text{adv}$ and $\theta_\text{ft}^\text{benign}$.
Following previous works~\cite{carlini2019secret, mireshghallah2022memorization}, we use the skew-normal distribution~\cite{o1976bayes} to model the perplexity distribution of secrets for efficiently approximating the exposure.
The precise exposure is upper-bounded by $\log_2|\mathcal{R}|$ when the target secret ranks the first among the whole set of possible secrets $\mathcal{R}$.
Thereby, the threshold $\log_2|\mathcal{R}|$ on the approximated exposure discriminates the case where a secret is only marginally the most likely or the case a secret is beyond the most likely.
A secret is only reliably extracted from the model with an exposure above the threshold~\cite{carlini2019secret}.
More specifically, we take secret as 10 digits in phone numbers and 10 English characters in email, thus derive $\log_2(10^{10})\approx 33$ and $\log_2(26^{10})\approx 47$ as the valid exposure threshold.
We can draw the following conclusion from \Cref{fig:targeted}.

\textbf{RQ8-Response}:
\textit{PreCurious} can use sanitization text to expose originally safe secrets even when scrubbing is perfect.

\section{Related Work}\label{chap:related}
We discuss the most related attacks and privacy risk amplification.

\textbf{Membership Inference Attack.} MIA in machine learning  context~\cite{shokri2017membership, ye2022enhanced} aims to predict whether a given sample is involved in training.
Considering the inefficiency of LLM training, we focus on threshold-based MIA as it is more practical than attack-model-based MIA~\cite{shokri2017membership, nasr2019comprehensive, choquette2021label, li2021membership}.
The key idea of threshold-based MIA is formalizing a hypothesis test with the posterior distribution assumptions about the model parameters~\cite{long2018understanding, carlini2022membership, ye2022enhanced}, by observing the signals from loss value~\cite{yeom2018privacy} or the loss calibrated by other models or samples~\cite{sablayrolles2019white, watson2021importance, long2020pragmatic, mireshghallah2022memorization, carlini2022membership, mattern2023membership}.
Our evaluations integrate both conventional loss signal~\cite{yeom2018privacy} and the state-of-the-art reference-model calibrated signal~\cite{mireshghallah2022memorization, ye2022enhanced, carlini2022membership} without retraining or multiple queries for each sample for a practical adversarial capability assumption.

\textbf{Data Extraction.}
Instead of extracting artificial canaries~\cite{carlini2019secret}, a previous work~\cite{carlini2021extracting} formulates the paradigm of extracting verbatim subsequence from the pre-training dataset of GPT-2 by filtering and ranking generated samples.
We evaluate the verbatim extraction on real secrets under this paradigm.

\textbf{Privacy Risk Amplification.}
The key idea of privacy risk amplification is to manipulate model or data integrity for more privacy leakage, as in representative works listed in \Cref{tab:related_work}. 
Prior works~\cite{tramer2022truth, chen2022amplifying, mahloujifar2022property} investigate the privacy risk amplification via data poisoning, which requires the control of the training dataset.
Recent work~\cite{tian2023manipulating} attempts to enlarge the property inference effect by manipulating the pre-trained encoder for image classification.
Our attack does not require control over the target training dataset and aims to plant a privacy backdoor in pre-trained model for amplifying general privacy risks in LLMs.
Concurrent works~\cite{feng2024privacy, wen2024privacy} also introduce privacy backdoors for pre-trained models, but \cite{feng2024privacy} is not comparable to ours as they focus on classification task and mainly assume stronger capabilities of white-box and architecture modification.
The other attack~\cite{wen2024privacy} is close to our basic version.
Our advanced strategies further consider random PEFT initialization and early-stopping performed by the victim.

\begin{table}[h]
\caption{Comparison with related works that manipulate integrity for privacy risk amplification. 
Manipulate: \protect\PartCircle{2}{2}/\protect\PartCircle{1}{2}/\protect\PartCircle{0}{2} represents manipulating model parameters/model/training data; 
PEFT: \protect\PartCircle{2}{2}/\protect\PartCircle{1}{2}/\protect\PartCircle{0}{2} represents no/evaluated/evaluated and investigated.
Case II: whether considering comparison cases when the fine-tuner applies early stopping.
Stealthy: whether considering stealthiness control.
}
\label{tab:related_work}
\resizebox{\columnwidth}{!}{%
\begin{tabular}{c|cccccc}
\toprule[1.2pt]
\textbf{Method}                                          & \textbf{Attacker's Goal}                                                                      & \textbf{Victim's Task}                                   & \textbf{Manipulate}                           & \textbf{Case II}                              & \textbf{Stealthy}                            & \textbf{PEFT}              \\
\midrule[1.2pt]
\cite{chen2022amplifying}      & MIA                                      & Discriminative   & \PartCircle{0}{2} & N/A                                  & yes & \PartCircle{2}{2} \\
\cite{mahloujifar2022property} & Property inference                                      & Discriminative & \PartCircle{0}{2} & N/A                                  & no & \PartCircle{2}{2} \\
\cite{tramer2022truth}         & MIA+Extraction & Generative   & \PartCircle{0}{2} & N/A                                  & no & \PartCircle{2}{2} \\
\midrule[1.2pt]
\cite{tian2023manipulating}    & Property inference                                      & Discriminative   & \PartCircle{2}{2} & N/A                                  & yes & \PartCircle{2}{2} \\
\cite{feng2024privacy}         & Reconstruction                                      & Discriminative  & \PartCircle{1}{2} & N/A                                  & no & \PartCircle{2}{2} \\
\cite{wen2024privacy}          & MIA                                      & Generative   & \PartCircle{2}{2} & no & yes & \PartCircle{1}{2} \\
\midrule[0.2pt]
Ours                                            & MIA+Extraction & Generative & \PartCircle{2}{2} & yes & yes & \PartCircle{0}{1} \\
\bottomrule[1.2pt]
\end{tabular}
}
\end{table}

\section{Discussion}

\textbf{Countermeasures.} We now discuss the countermeasures to PreCurious for the wide range of users and regularization designers.

\textit{Be careful to download models from unknown sources.}
The amplified risk from \textit{PreCurious} justifies the importance of model integrity in pre-training and fine-tuning pipeline.
Therefore, we recommend that fine-tuners download pre-trained models from trusted sources rather than from anonymous users on open-source platforms.
Users should check the download link and be aware when automatic library management tools upgrade to higher version packages.

\textit{Be careful when following fine-tuning instructions.}
With the rapid development of language models, users with different backgrounds can get started on building their models easily by following tutorials from the community.
However, the success of \textit{PreCurious} reveals additional side information that can be exploited by the adversary to infer private information.
Users should not rely heavily on common settings shared in a tutorial, but instead be aware of the training dynamics in fine-tuning (e.g., epochs, stopping criteria, PEFT choices), even as the validation loss continues to decrease.

\textit{Be careful on auditing risks even under defense.}
\textit{PreCurious} demonstrates that  regularization defense, DP fine-tuning, and deduplication are not perfect.
For example, DP even with a strict budget cannot lead to a random guess attack under \textit{PreCurious}; deduplication fails when attackers can approximate the deduplicated text in MIA, or when \textit{PreCurious-lagging} implicitly increases the number of repetitions for all samples.
Thus, we suggest that users remain vigilant and audit the privacy dynamics during fine-tuning closely~\cite{carlini2022membership, mireshghallah2022memorization, carlini2019secret} even when reasonable defenses are applied.

\textit{Be careful to share sanitized text by masking PII.}
\textit{PreCurious} demonstrates the feasibility of increasing the risk of secret exposure by using a public sanitized dataset to improve the auxiliary knowledge.
Thus, we claim that unless we can ensure that sensitive information is removed for each future training, it is not safe to publish sanitized datasets, even if the sensitive secrets are masked or replaced, which is important when researchers in high-stakes domains publish benchmark datasets.

\textbf{Implications for future works.}
A recent work~\cite{ye2023initialization} investigates the influence of model initialization on the worst-case privacy risk scales with the gradient difference on neighboring datasets and the iterations.
\textit{PreCuious} fills the gap between the theoretical discussion on model initialization from scratch and the practical use of pre-trained LMs and PEFT technique from an average case perspective.
It is interesting for future work to improve the theoretical understanding of worst-case privacy when applying model efficiency techniques, as well as to exploit other side information to explore potential vulnerabilities for evaluating existing defenses.

From \textit{PreCurious}, we note that memorization-based privacy backdoors on either accelerating or lagging direction should be coupled with the stopping criteria to derive the final risk amplification effect.
Since there is no privacy attack considered to improve risks when victims perform early stopping, we bring new perspectives for future attacks and defenses under this realistic scenario.
In addition, \textit{PreCurious} reveals the vulnerability and identifies corner cases of existing defenses, providing a critical call for stronger defenses.

\section{Conclusion}
In this paper, we introduced \textit{PreCurious}, a novel privacy risk amplification framework that increases the privacy risk of fine-tuning dataset by manipulating the pre-trained model's memorization level and releasing a crafted model, 
showing the importance of model integrity from the privacy lens.
We are among the first to investigate privacy backdoors, throughly exploring cases of PEFT and early-stopping by leveraging the side information in fine-tuning guideline.
Our findings show that \textit{PreCurious} breaks up the privacy-invulnerability property for PEFT, and common-sense defenses are possible to be subverted.
Our work takes the step to understand the interplay between model memorization, efficiency and privacy risks, while also raises an interesting perspective to break up privacy-utility trade-off.
This research is a critical call to action, urging the community to improve safeguards and reevaluate the security protocols around the use of pre-trained models, particularly those sourced from unverified platforms.

\begin{acks}
We would like to thank reviewers for their constructive comments and efforts in improving our paper and artifacts.
This work was supported in part by  
National Science Foundation grants (CNS-2125530, CNS-2124104, IIS-2302968, CNS-2350333), 
National Institutes of Health grants (R01LM013712, R01ES033241),
JSPS KAKENHI JP23K24851, JST PRESTO JPMJPR23P5, and JST CREST JPMJCR21M2.
\end{acks}

\bibliographystyle{ACM-Reference-Format}
\bibliography{src/ref_att}

\appendices
\section{Experimental Setup and More Results}
Our experiments were conducted on an Ubuntu 20.04.6 system with 8 NVIDIA Quadro RTX 8000 GPUs.
The source code and other artifacts have been made available~\footnote{\url{https://github.com/Emory-AIMS/PreCurious}}.

\begin{table}[!h]
\centering
\caption{MIA effectiveness under DP-SGD defense on PTB dataset with AdapterFT.}
\label{tab:defense_dp_ptb_adapter_APP}
\resizebox{0.78\columnwidth}{!}{%
\begin{tabular}{c|c|ccc|c}
\toprule[1pt]
$\epsilon$ & MIA metric & TPR@0.01FPR & TPR@0.1FPR & AUC     & Val-PPL    \\
\midrule[1pt]
0.05 & Benign     & \textbf{2.29\%} & 10.03\%          & 51.99\%          & 73.64          \\
0.05 & PreCurious & 0.86\%          & \textbf{12.89\%} & \textbf{53.53\%} & \textbf{27.64} \\
0.5  & Benign     & \textbf{1.72\%} & 10.03\%          & 52.03\%          & 70.41          \\
0.5  & PreCurious & 1.43\%          & \textbf{13.47\%} & \textbf{55.09\%} & \textbf{26.42} \\
1    & Benign     & \textbf{1.72\%} & 10.03\%          & 52.05\%          & 68.61          \\
1    & PreCurious & 0.86\%          & \textbf{14.04}\%          & \textbf{54.84\%} & \textbf{25.94} \\
2    & Benign     & \textbf{1.72\%} & 9.74\%           & 52.01\%          & 66.59          \\
2    & PreCurious & 1.15\%          & \textbf{14.33\%} & \textbf{54.58\%} & \textbf{25.47} \\
\bottomrule[1pt]
\end{tabular}
}
\end{table}

\begin{table}[!h]
\centering
\caption{Untargeted $p_\text{ext}\uparrow$ on PTB with Adapter-FT.}\label{tab:un_ptb_adapter_APP}
\resizebox{0.76\columnwidth}{!}{%
\begin{tabular}{c|c|cccc}
\toprule
$\epsilon$ & Pre-trained model & \multicolumn{4}{c}{Subsequence Length} \\
\cmidrule{3-6}  %
           &        (w/ or w/o Ref)           & 2       & 5       & 10      & 50      \\
\midrule
0.05 & PreCurious w/ Ref  & 91.78\% & 57.85\% & 39.43\% & 18.10\% \\
0.05 & PreCurious w/o Ref & 56.95\% & 49.65\% & 37.10\% & 19.80\% \\
0.05 & Benign w/ Ref      & 65.68\% & 39.20\% & 37.08\% & 20.94\% \\
0.05 & Benign w/o Ref     & 46.67\% & 41.34\% & 36.84\% & 18.33\% \\
8    & PreCurious w/ Ref  & 92.88\% & 58.81\% & 39.04\% & 18.67\% \\
8    & PreCurious w/o Ref & 65.43\% & 57.67\% & 37.13\% & 19.92\% \\
8    & Benign w/ Ref      & 62.53\% & 39.21\% & 37.20\% & 21.32\% \\
8    & Benign w/o Ref     & 44.11\% & 39.20\% & 36.88\% & 18.84\% \\
\bottomrule
\end{tabular}
}
\end{table}

\end{document}